%% file: Auto-Formula.tex
\documentclass[acmsmall]{acmart}
\usepackage{etoolbox}

\usepackage{comment}
\usepackage{xspace}
\usepackage{xr-hyper}
\usepackage{url}
\usepackage{graphicx}
\usepackage{balance}  
\usepackage{xcolor}
\usepackage{amsmath}
\usepackage{enumitem}
\usepackage{subfigure}
\usepackage{cleveref}
\usepackage{url}
\usepackage{multirow}
\usepackage{amsmath}
\usepackage{float}
\usepackage[ruled,linesnumbered]{algorithm2e}

\newtoggle{fullversion}
 \toggletrue{fullversion}

\NewDocumentCommand{\fanj}{mO{}}{\textcolor{red}{\textsuperscript{\textit{fanj}}\textsf{\textbf{\small[#1]}}}}

\newcommand{\stitle}[1]{\vspace{1ex}\noindent{\bf #1}}
\newcommand{\etitle}[1]{\vspace{1mm}\noindent{\underline{\em #1}}}
\newcommand{\sys}{Auto-Formula\xspace}

\newcommand{\code}[1]{{\small \texttt{#1}}}

\newcommand{\add}[1]{\textcolor{black}{#1}}

\makeatletter
\DeclareRobustCommand*\cal{\@fontswitch\relax\mathcal}
\makeatother

\newcounter{definition}
\newenvironment{definition}[1][]{\refstepcounter{definition}\par\smallskip\textsc{Definition~\thedefinition.\ #1}}{\smallskip}

\newcounter{example}
\newenvironment{example}[1][]{\refstepcounter{example}\par\smallskip\textsc{Example~\theexample.\ #1}}{\smallskip}

\AtBeginDocument{%
  \providecommand\BibTeX{{%
    \normalfont B\kern-0.5em{\scshape i\kern-0.25em b}\kern-0.8em\TeX}}}

\acmConference[SIGMOD '20]{2020 International Conference on Management of Data}{June 14--19, 2020}{Portland, OR}

\usepackage{xr}
\makeatletter

\newcommand*{\addFileDependency}[1]{
\typeout{(#1)}
\@addtofilelist{#1}
\IfFileExists{#1}{}{\typeout{No file #1.}}
}\makeatother

\newcommand*{\myexternaldocument}[1]{%
\externaldocument{#1}%
\addFileDependency{#1.tex}%
\addFileDependency{#1.aux}%
}

\myexternaldocument{cover}

\begin{document}
\pagestyle{plain} 

\title{\sys: Recommend Formulas in Spreadsheets using Contrastive Learning for Table Representations}




\author{Sibei Chen}
\authornote{Part of work done while at Microsoft.}
\affiliation{%
  \institution{Renmin University of China}
  }
\orcid{0009-0001-5331-5829}
\email{sibei@ruc.edu.cn}

\author{Yeye He}
\affiliation{%
  \institution{Microsoft Research}
  }
\orcid{0000-0003-2824-5299}
\email{yeyehe@microsoft.com}

\author{Weiwei Cui}
\affiliation{%
  \institution{Microsoft Research}
  }
\orcid{0000-0003-0870-7628}
\email{weiweicu@microsoft.com}

\author{Ju Fan}
\affiliation{%
  \institution{Renmin University of China}
  }
\orcid{0000-0003-4729-9903}
\email{fanj@ruc.edu.cn}

\author{Song Ge}
\affiliation{%
  \institution{Microsoft Research}
  }
\orcid{0000-0003-2178-3086}
\email{songge@microsoft.com}

\author{Haidong Zhang}
\affiliation{%
  \institution{Microsoft Research}
  }
\orcid{0000-0001-7530-9553}
\email{haizhang@microsoft.com}

\author{Dongmei Zhang}
\affiliation{%
  \institution{Microsoft Research}
  }
\orcid{0000-0002-9230-2799}
\email{dongmeiz@microsoft.com}

\author{Surajit Chaudhuri}
\affiliation{%
  \institution{Microsoft Research}
  }
\orcid{0000-0001-8252-5270}
\email{surajitc@microsoft.com}


\setcopyright{acmlicensed}
\acmJournal{PACMMOD}
\acmYear{2024} \acmVolume{2} \acmNumber{3 (SIGMOD)} \acmArticle{122} \acmMonth{6}\acmDOI{10.1145/3654925}



\received{October 2023}
\received[revised]{January 2024}
\received[accepted]{February 2024}


\begin{abstract}
Spreadsheets are widely recognized as the most popular end-user programming tools, which blend the power of formula-based computation, with an intuitive table-based interface. Today, spreadsheets are used by billions of users to manipulate tables, most of whom are neither database experts nor professional programmers. 

Despite the success of spreadsheets, authoring complex formulas remains challenging, as non-technical users need to look up and understand non-trivial formula syntax. 
To address this pain point, we leverage the observation that there is often an abundance of similar-looking spreadsheets in the same organization, which not only have similar data, but also share similar computation logic encoded as formulas.  We develop an \sys system that can accurately predict formulas that users want to author in a target spreadsheet cell, by learning and adapting formulas that already exist in similar spreadsheets, using contrastive-learning techniques inspired by ``similar-face recognition'' from compute vision.
Extensive evaluations on over 2K test formulas extracted from real enterprise spreadsheets show the effectiveness of \sys over alternatives. Our benchmark data is available at \url{https://github.com/microsoft/Auto-Formula} to facilitate future research.
\end{abstract}

\maketitle

\input{intro.tex}
\input{related.tex}
\input{problem.tex}
\input{framework}

\input{experiment.tex}

\input{conclusions.tex} 

\begin{acks}
    We thank three anonymous reviewers for their constructive feedback, as well as the helpful feedback from Microsoft's Excel Formula AI team.
    
	This work was partly supported by the NSF of China (62122090 and 62072461), the Beijing Natural Science Foundation (L222006), the Research Funds of Renmin University of China, and the Outstanding Innovative Talents Cultivation Funded Programs 2024 of Renmin University of China.
\end{acks}

\break
\bibliographystyle{ACM-Reference-Format}
\bibliography{Auto-Formula}

\end{document}

%% file: intro.tex
\section{Introduction}
\label{sec:intro}

Spreadsheets, such as those in Microsoft Excel and Google Sheets, are commonly recognized as the most popular end-user programming tools to manipulate tabular data~\cite{hermans2016spreadsheets, erwig2009software}. 
The intuitive spreadsheet interface combines the power of formula calculations, with the ability to visually inspect tables, and is widely used by billions of non-technical users (e.g., average enterprise users), who are neither database experts nor professional programmers.

\begin{figure*}[t]
    \centering   \includegraphics[width=1.02\textwidth]{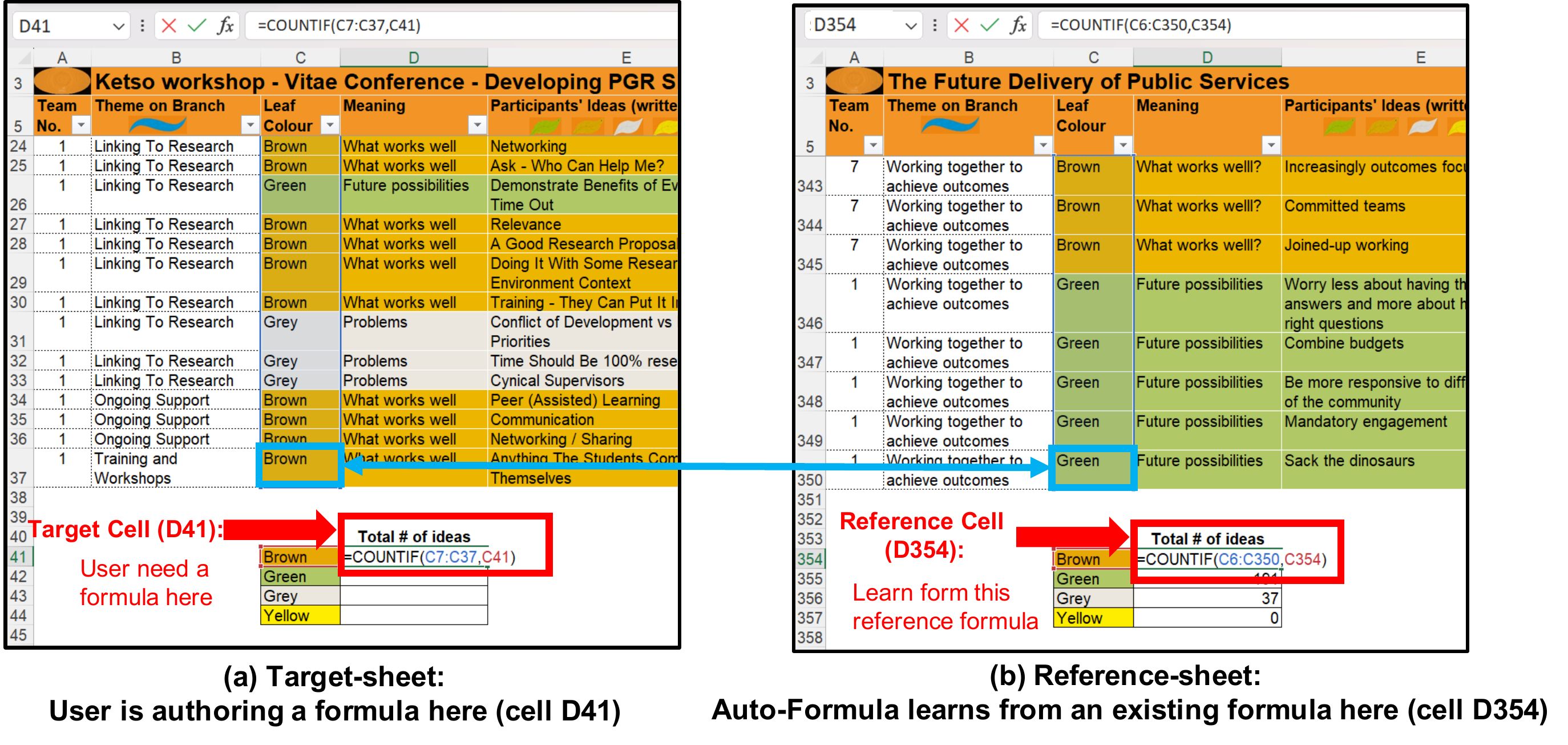}
    \caption{A pair of example spreadsheets that are ``similar-sheets'', with similar style and color that are easy for humans to spot (though the two have different data content, and different number of rows/columns). (Left) Target-sheet: a user is trying to author a new formula on this target-spreadsheet, in the target-cell (D41, left), where the ground-truth formula should be ``\code{=COUNTIF(C7:C37,C41)}''; (Right) \sys learns-to-retrieve an existing sheet on the right as a “similar-sheet”, and adapts an existing formula from cell D354 (because D354's surrounding region is similar to that of D41 on the left), to the target-cell (D41, left), by using the same formulate-template but changing the parameters based on the local context of the target cell (D41, left). }
    \label{fig:formula-ex}
 \vspace{-3mm}
\end{figure*}

\textbf{User pain point: creating formulas}. Despite the popularity of spreadsheets, creating formulas in spreadsheets remains a key pain point for non-technical users -- there is a long line of prior studies that show it is both 
challenging~\cite{spreadsheet-study-1, spreadsheet-study-2, spreadsheet_study_3, spreadsheet-study-4, zhao2024nl2formula}) 
and error-prone (\cite{spreadsheet-error-1, spreadsheet-error-2, spreadsheet-error-3}) for users to create formulas from scratch in spreadsheets. In order to create any non-trivial formula  in spreadsheets (beyond the very simple \code{AVG()} and \code{SUM()}), users would first need to identify relevant functions, and then  understand the function syntax by reading documentations, which is close to what is expected from  professional developers and not trivial for non-technical enterprise users.

Figure~\ref{fig:formula-ex}(a) shows a real example spreadsheet, where a user is trying to author a new formula in cell \code{D41}, with the intent to count the number of occurrences of the value displayed to the left (``\code{Brown}'', in cell \code{C41}), within the column of data from cell \code{C7} to \code{C37}. The ground-truth formula is therefore ``\code{$=$COUNTIF(C7:C37,C41)}'', which however is non-trivial for users to write, as an average spreadsheet user is typically not familiar with the \code{COUNTIF()} function and its syntax.  Note that this simple formula has uses only one \code{COUNTIF()} function -- today's spreadsheet software supports hundreds of
functions (674 functions for Excel \cite{efl}, 
and 512 for Google Sheets~\cite{gfl} as of April 2024). 
Furthermore, our profiling of real spreadsheets shows that over $43\%$ of formulas use multiple functions, and over $59\%$ of formulas have multiple parameters, 
each of which needs to be programmed correctly, and then stringed together appropriately, making the task of authoring formulas clearly challenging.




Given the aforementioned challenges, it is no surprise that lots of users find it hard to program formulas, as evidenced by large numbers of such questions from user forums -- for example, a single Excel user forum in~\cite{efc} shows over 20K user questions tagged as ``formulas and functions'', which underscores the scale of the challenges faced by users in authoring formulas for spreadsheets.

It would clearly be useful if we can help users to author formulas, by predicting the desired formulas in a given spreadsheet cell, which can then be shown to users as ``\emph{recommended formulas}'' for users to verify and choose from.

\textbf{Prior approach: using natural language context}. To our knowledge, there is only one recent work from Google that attacks the same formula-recommendation problem, called SpreadsheetCoder~\cite{spreadsheetcoder} 
that is studied in the context of Google Sheets~\cite{gsbs}, where the authors propose to predict formula in a user-specified target cell,  based on its surrounding Natural Language (NL) context -- intuitively, if the row label or column header of the target cell is ``Total'' or ``Aggregate'', then the desired formula will likely involve the function \code{SUM()} for the corresponding row or column.

Although this is clearly an interesting approach, we find it insufficient in our tests, especially when the target formula is complex. For example, the target formula shown in Figure~\ref{fig:formula-ex} uses the function \code{COUNTIF()} with two parameters, which is hard to infer from the natural language context alone, because even as humans, we may not be able to correctly guess this \code{COUNTIF()} formula by looking at the surrounding NL-context alone (without knowing the actual intent of the user for a formula in Cell \code{D41}). In our tests on real-world formulas extracted from real spreadsheets, we find this prior approach in~\cite{spreadsheetcoder} to have low accuracy, and can only predict relatively simple short formulas. 

\textbf{\sys: Leveraging ``similar-sheets''}. Given the difficulty of predicting formulas based solely on the surrounding NL-context, in this work, we propose an alternative approach that predicts formulas based on ``similar-sheets''. 

\underline{Intuition for how to predict accurately}. Our key observation is that in the same organization, a significant fraction of spreadsheets have similar-looking counterparts, like depicted in the pair of examples in Figure~\ref{fig:formula-ex}. Such ``similar-sheets'' share similar patterns and content, both syntactically (e.g., in terms of color schemes and fonts), and semantically (in terms of data content and formula-logic). We find such similar-sheets often represent different subsets of data in practice -- e.g., financial statements for different time periods, or sales reports for different geo locations. 

In order to quantify the prevalence of ``similar-sheets'' in practice, we sample spreadsheets from  4 large organizations (Enron, PGE, TI, and Cisco, all crawled from public sources). We sample spreadsheets for a manual inspection, and found that about $40-90\%$ of spreadsheets have similar-sheet counterparts like in Figure~\ref{fig:formula-ex}, showing its ubiquity.

Our unique insight is that such similar-sheets may already contain similar formula logic, programmed by other users in the same organization, and can be leveraged to accurately predict a new formula that users want to author. For example, imagine that a user is trying to author a formula in \code{D41} of the spreadsheet in Figure~\ref{fig:formula-ex}(a), which we will refer to as the ``\emph{target-sheet}'', because \code{D41} is the ``\emph{target-cell}'' that users want to fill in.  If we can identify a similar-sheet of this target-sheet, shown in Figure~\ref{fig:formula-ex}(b), which we refer to as a ``\emph{reference-sheet}'', and pinpoint a relevant spreadsheet region centered around cell \code{D354}, which we call the ``\emph{reference-cell}'', we can actually leverage this existing formula already programmed in the reference-cell, to accurately predict the new formula that a user would need in the target-cell on the left. We emphasize that this approach of ``learning-from-similar-sheet'' is much more reliable than leveraging NL-context only, especially for complex formula-logic involving multiple functions that are hard even for humans to guess from the target-sheet alone.

\underline{Key technical challenges}.
In this work, we explore the new direction of ``formula-recommendation by similar-sheet'' that is not considered in the literature, where we face a number of important technical challenges.

First, we need to reliably identify ``similar-sheets'' from large numbers of existing spreadsheets in the same organization, which is challenging because two related sheets that may look apparent to human eyes (e.g., the two in Figure~\ref{fig:formula-ex}), can still have different content/design/color etc., and may be ``shifted'', with different numbers of rows and columns (e.g., the ``target-table'' in Figure~\ref{fig:formula-ex}(a) has only 37 rows, while the ``reference-sheet'' in Figure~\ref{fig:formula-ex}(b) has 350 rows), making a naive cell-by-cell comparison approach ineffective. 

Second, to make the matter worse, formula recommendation is an online ``user-in-the-loop'' experience with strict latency requirement (e.g., a couple seconds at most), yet organizations today have large amounts of spreadsheets, where the similar-sheets we aim to find are like ``a needle in a haystack'', which poses substantial efficiency and scalability challenges for this to be tractable.

Lastly, even after similar-sheets and similar-regions can be identified, we need to automatically rewrite/adapt an existing formula from the reference-cell (e.g., ``\code{=COUNTIF(C6:C350,C354)}'' in Figure~\ref{fig:formula-ex}(b)) into the new ``context'' of the target-sheet (e.g., ``\code{=COUNTIF(C7:C37,C41)'' in Figure~\ref{fig:formula-ex}(a)}, which has the same formula-template, but different parameters referencing different cells of the target-sheet). This is yet another challenge to address before an entire formula can be predicted correctly for users.

\underline{Similar-sheets as a ``face-recognition'' problem}. To predict formulas and find similar-sheets, we develop a novel approach that models spreadsheets (consisting of cells in a two-dimensional grid), as ``pictures'' in computer vision (consisting of pixels in a two-dimensional grid), so that our ``similar-sheet'' problem becomes analogous to the classical ``face recognition'' problem (e.g., identifying similar faces belonging to the same person), studied extensively in the computer-vision literature~\cite{facenet, cosface, sphereface}.

We build an \sys system for this problem, where we first employ weak supervision to automatically harvest large amounts of training data for ``similar-sheets'', from a large corpus of real spreadsheets. We then train model representation of spreadsheets using dense embedding vectors, based on a model architecture adapted from classical computer-vision (e.g., FaceNet and triplet loss~\cite{facenet}), but tailored specifically to our spreadsheet problem.  
Our system is shown to be effective when extensively evaluated using real spreadsheets. 


\textbf{Contributions}. We make the following contributions:

\begin{itemize}[noitemsep,topsep=0pt,leftmargin=*]
\item We develop an \sys system for the important problem of formula recommendation, using a novel approach that learns-to-adapt formulas from similar spreadsheet, which is substantially more accurate than existing methods from the literature and in commercial systems.
\item We propose two key primitives called ``similar-sheet'' and ``similar-region'', using representation of spreadsheets learned via contrastive-learning, that are inspired by ``face recognition'' in computer-vision. These primitives are crucial for formula recommendation and can be of independent interest in other spreadsheet applications.
\item We systematically evaluate different methods for the formula-recommendation problem, using large amounts of real formulas extracted from spreadsheets for the first time, which we crawl and compile from public sources, that we will release for future research. 
\end{itemize}

%% file: related.tex
\section{Related Work}
\label{sec:related}

\textbf{Spreadsheet formula suggestion.}  SpreadsheetCoder~\cite{spreadsheetcoder} developed in the context of Google Sheets, is the only recent work in the literature we are aware of  that also studies the formula-recommendation problem. It uses the natural language contexts in the surrounding cells to predict the desired formula in a target cell (e.g., if column header says ``total'' or ``sum'', then a \code{SUM()} function will likely be used).  In our experiments, we find this approach has low accuracy, especially for complex formulas with multiple functions and parameters (e.g., Figure~\ref{fig:formula-ex}).


There is a related, but different type of coding-assistant, known as NL-to-code or semantic parsing~\cite{pasupat2015compositional, berant2014semantic, zhao2024nl2formula}, where users would type a natural-language query (e.g., ``SUM of sales for FY23'', or ``Count the number of Brown values in Column C'' for the formula in Figure~\ref{fig:formula-ex}), for the system to generate the desired formula/code. 
In contrast, in our formula-recommendation problem 
(the problem studied in both SpreadsheetCoder and \sys), predictions are made in a ``contextual'' manner, 
and without needing users to actually type in a natural language query, which provides a friction-less experience as demos 
like~\cite{gsbs, google_sheets_demo_backup} would show.

\textbf{Similar tables and spreadsheets.} 
The problem of discovering similar table is studied in the context of data lakes, e.g., union-table search~\cite{UnionTable}, where tables with similar schemas are discovered.  Compared to tables, the spreadsheets we study are richer (with text and non-text features), and more complex with no known table-boundaries (e.g., a spreadsheet may contain multiple tables, with a target formula outside of any table like in Figure~\ref{fig:formula-ex}), which renders existing techniques not directly applicable. 

In terms of similar spreadsheet discovery, Mondrian~\cite{mondrian} is the only prior work we are aware of, where they pioneered an interesting task of layout detection by clustering spreadsheets. Mondrian models regions in a spreadsheet as graph nodes, and uses a hand-crafted function to measure similarity. We show in our experiments that our learned spreadsheet representation is not only more accurate, but also orders of magnitude more efficient on large spreadsheet corpus, because our embedding-based search leverages recent advances in approximate nearest neighbor (ANN) search~\cite{faiss, ann-1, ann-2}. 

\textbf{Face recognition in computer vision.} The ``similar-sheet'' and spreadsheet representation technique we develop in this work, is inspired by ``face recognition'' in computer vision~\cite{cosface, facenet, sphereface},  where  accurately finding ``similar faces'' (belonging to the same person)  from an ocean of faces in a database is the key challenge. If we view spreadsheets as images of faces, then spreadsheet-cells are naturally like image-pixels, but instead of using RGB channels to represent pixels, we use each cell's syntactic (color, font, size, etc.) and semantic features (content embedding). We design architectures inspired by computer vision but tailored specifically to our problem on spreadsheet tables, which we show are effective in the table domain.

\textbf{Other forms of suggestions in programming contexts.} In the general programming context, IntelliSense in Visual Studio~\cite{intellisense} and GitHub Copilot~\cite{github_copilot} are some of the well-known examples that provide contextualized auto-complete in IDEs, which are similar in spirit to \sys, as both provide contextualized auto-complete-like recommendations, without requiring users to issue natural-language queries. However, these systems target general programming environments that focus on ``code'', without considering ``tables''. This is unlike \sys that is designed for spreadsheets, where ``tables'' is a first class citizen, and the interaction between code/table is the central focus for spreadsheet-formula recommendations.

In other data-table and data-pipeline related contexts, Auto-Suggest~\cite{yan2020auto} proposes to predict operator parameters in data pipelines and notebooks, using table-level features that can tailor to different operators. EDAssistant~\cite{li2023edassistant} can suggest EDA operations in the context of exploratory data analysis pipelines and notebooks. Auto-Pipeline~\cite{yang2021auto} and Auto-Pandas~\cite{bavishi2019autopandas} learn to predict multi-step pipelines using many (input tables, target output) pairs.  HAIPipe~\cite{chen2023haipipe} and Learn2Clean~\cite{berti2019learn2clean} can learn to select suitable operations in ML pipelines. 

In spreadsheet environments, Program-by-Example systems such as FlashFill and TDE~\cite{auto-transform, tde, gulwani2011automating, he2018transform, zhu2017auto, gulwani2016programming, foofah} ask users to provide input/output examples in order to synthesize programs/formulas, which is another class of popular coding assistants. 
Auto-Tables~\cite{auto-tables} further learns-to-suggest transformations without examples, based on the characteristics of input tables.

In the context of SQL, there are a number of additional systems that can specifically suggest SQL queries or templates, such as QueRIE~\cite{akbarnejad2010sql}, SnipSuggest~\cite{khoussainova2010snipsuggest}, SQLSugg~\cite{fan2011interactive}, and Seq~\cite{lai2023workload}, etc.

All of these are orthogonal to spreadsheet formula-prediction, where code and tables are tightly blended in the same spreadsheet grid, which is the unique characteristics that we leverage to recommend formulas in this work (all without using other forms of input, such as natural language queries, input/output examples, or query session information).

%% file: problem.tex
\vspace{-2mm}
\section{Preliminaries}
\label{sec:prelim}
\subsection{Spreadsheets and spreadsheet tables}


Spreadsheets are widely used by billions of end-users to store and analyze data, in places like Microsoft Excel and Google Sheets. Similar to relational tables in databases, spreadsheets also store tabular data in a row-and-column format like in Figure~\ref{fig:formula-ex}. However, there are important differences between the two, making spreadsheets more challenging to deal with.

\etitle{No clear table boundary and structure}. Spreadsheets are two-dimensional but allow flexible organization of data (table content), metadata (captions, hierarchical headers, etc.), and free-form texts, organized in ad-hoc ways (e.g., a traditional ``database column'' may be stored in the horizontal direction). The fact that spreadsheets lack explicit machine-understandable structures makes them particularly challenging. 

\etitle{Mixed data and code/formula}. Spreadsheets blend both data and code/formulas, in the same two-dimensional grid, where formulas are programmed at the granularity of individual cells (e.g., the formula in a cell can often be different from the formula in the neighboring cell). This rich interplay between data and code is unique yet powerful, which poses new research challenges for our community.

    
\etitle{Rich non-textual styles}. Unlike database tables or CSV/JSON tables, where each cell is just a string, spreadsheet come with rich non-text styles (e.g., font, color, borders, size, etc., like in Figure~\ref{fig:formula-ex}), which are meant to enhance readability for humans. 
Such visual features offer clues for humans, and are equally important for our  computer-vision-inspired algorithms.

\stitle{Similar sheets:} We observe that a large fraction (40\%-90\% based on our studies) of spreadsheets created in the same organization often exhibit a high degree of similarity, often with similar data/formula, and serve similar purposes (e.g., the financial statements for different time periods, or the sales report for different geo-locations), like shown in Figure~\ref{fig:formula-ex}. This is a key property that we leverage to accurately predict complex formulas. 


\subsection{Excel Formulas}

In spreadsheets, a formula is a user-programmed expression in a cell, which consists of: (1) functions such as \code{SUM()} and \code{COUNTIF()}, from hundreds of such options~\cite{gf, ef}, 
and (2) parameters to these functions, which are often cell-locations like ``\code{B5}'' that refer to data in other cells. 



\stitle{Formula templates.} Given a concrete formula $F$, we can write it as $F = \overline{F}(R)$, where $\overline{F}$ is a ``\emph{formulate template}'' that consists of the functions and AST (abstract syntax tree) structure of the formula $F$.  Such a formulate template $\overline{F}$ does not have specific parameters, instead has ``holes'' that need to be filled in, to produce a concrete formula. For instance, in Figure~\ref{fig:formula-ex}, the formula $\code{=COUNTIF(C7:C37,C41)}$ has a formula template of $\code{=COUNTIF(\_:\_,\_)}$, with three placeholders for parameters.




\stitle{Parameter cells.} Parameter cells, written as $R$, reference to data in other cell locations, e.g., $\code{C41}$, and can be used as parameters to instantiate a formula template $\overline{F}$ into a concrete formula.
Because parameter cells reference the (dynamic) content of other locations, they
are similar to ``variables'' used in programming context. 

Note that to work on two-dimensional tables, we can also have ``parameter cell ranges'', which refer to multiple cells in a continuous range (e.g., a row or column), such as $\code{C7:C37}$ shown in  Figure~\ref{fig:formula-ex}  (it specifies a column with 31 cells, in which a ``count'' operation needs to be performed).



\subsection{Problem: Formula Recommendation}
\label{sec:problem}


We will first introduce \emph{target sheet} and \emph{target cell}, before defining our ``formula recommendation'' problem.

\begin{itemize}
    \item \textbf{Target sheet}, denoted by $S_{T}$, is a spreadsheet currently edited by users.
    \item \textbf{Target cell}, denoted by $C_{T}$, is a cell in which the user wants to create a formula.
\end{itemize}

\begin{definition}
    \textit{[Formula Recommendation].} For a given target sheet $S_{T}$ and a target cell $C_{T} \in S_{T}$, formula-recommendation is the problem of predicting the desired formula $F_{T}$ in $C_{T}$, based on the context of the sheet $S_{T}$. 
\end{definition}

We emphasize that in our problem, a predicted formula is correct only if both its formula template $\overline{F}_{T}$, and parameter cells $R_{T}$, can
\emph{completely} match the ground-truth (the actual formula entered by users in the spreadsheet). For example, to predict the example in Figure~\ref{fig:formula-ex} correctly, the algorithm needs to predict both the formula template $=\code{COUNTIF(\_:\_,\_)}$, and each parameter cell correctly, which are $\code{C7}$, $\code{C37}$ and $\code{C41}$, respectively.

Furthermore, in order to be general, we specifically include any valid spreadsheet formulas that can be parsed by a formula parser. The formulas considered in our problem can therefore be arbitrarily complex, with functions, cells, cell ranges, constants, etc., and and can be defined in a recursive manner.



    


%% file: framework.tex
\section{\sys by similar-sheets}
\label{sec:method}



We propose to learn-to-predict formulas, leveraging existing formulas that are already authored on similar spreadsheets.

\subsection{Intuition and Architecture Overview} \label{sec:intuition}
In this section, we will start by giving an intuitive explanation of how \sys work.

Recall that unlike prior work~\cite{spreadsheetcoder}, we propose to predict formulas leveraging an existing corpus of spreadsheets (e.g., from the same organization), denoted by $\mathbf{S}$.
At a high level, our approach works in three intuitive steps: 
\begin{itemize}[noitemsep,topsep=0pt,leftmargin=*]
\item[] \underline{(S1) Search reference-sheets (by similar-sheet)}: In this first step, we identify one or more \emph{reference-sheets}, denoted by $S_R \in \mathbf{S}$, that are similar to our target sheet $S_T$, using a new learned primitive we develop called ``\emph{similar-sheet}'';
\item[] \underline{(S2) Search reference-formula (by similar-region)}: within each reference sheet $S_R$, we identify a \emph{reference-cell} $C_R \in S_R$, whose ``local spreadsheet regions'' look similar to the ``local region'' around the target-cell $C_T$, using a ``\emph{similar-region}'' primitive we develop. If $C_R$ contains a formula $F_R$, this is a promising \emph{reference-formula}.
\item[] \underline{(S3) Search parameter-cells (by similar-region)}: 
 Next, we need to adapt the reference-formula $F_R$, into the local context of the target-sheet $S_T$, by changing the  parameter-cells of $F_R$. We again use the ``similar-region'' primitive to learn-to-adapt these parameters from $S_R$ to $S_T$.
\end{itemize}

\noindent We explain this prediction process using an example below.

\begin{example}
\label{ex:3steps}
    We revisit our running example in Figure~\ref{fig:formula-ex}. Recall that a user is trying to create a formula in our target-cell \code{D41}, from the left sheet (Figure~\ref{fig:formula-ex}(a)), which is our target-sheet $S_T$. 
    
    In the first step \underline{(S1) search reference-sheets}, we search in our spreadsheet corpus $\mathbf{S}$, to find reference-sheets that look similar to $S_T$, using the ``similar-sheet'' primitive. The sheet shown in Figure~\ref{fig:formula-ex}(b) is one such  reference-sheet $S_R$. 
    
    Next, in  \underline{(S2) search reference-formula}, from the reference-sheet in Figure~\ref{fig:formula-ex}(b), we search for a region that looks similar to the region around our target-cell \code{D41} in Figure~\ref{fig:formula-ex}(a), using a primitive we call ``similar-region''.  The region around the cell \code{D354} in Figure~\ref{fig:formula-ex}(b) looks very similar to the region around \code{D41}, so we use the formula contained in \code{D354}, ``\code{=COUNTIF(C6:C350,C354)}'', as our reference-formula $F_R$ in the next step.

    Finally, in \underline{(S3) search parameter-cells},  we will use the reference-formula $F_R$ to predict the formula in the target-cell $C_T$ (\code{D41} of Figure~\ref{fig:formula-ex}(a)).  Specifically, we utilize the  formula template of $F_R$, in this case \code{COUNTIF(\_:\_,\_)}, and we will try to fill in the parameters appropriately based on the local context of the target sheet $S_T$ (Figure~\ref{fig:formula-ex}(a)), because using the original parameters \code{C6}, \code{C350} and \code{C354} in $F_R$, taken from the reference-sheet $S_R$ in Figure~\ref{fig:formula-ex}(b), will likely be incorrect on the target-sheet in Figure~\ref{fig:formula-ex}(a). Instead, we use the regions centered around \code{C6}, \code{C350} and \code{C354} in  Figure~\ref{fig:formula-ex}(b), to look for similar-looking regions in the target-sheet Figure~\ref{fig:formula-ex}(a). We find the regions around  \code{C7}, \code{C37} and \code{C41} of  Figure~\ref{fig:formula-ex}(a) to be the most similar to \code{C6}, \code{C350} and \code{C354} of Figure~\ref{fig:formula-ex}(b), respectively, also using the  ``similar-region'' primitive. We can now fill the formula-template using these cells, to produce \code{COUNTIF(C7:C37,C41)}, which is the correct formula that the user wants in the target cell \code{D41}.

\end{example}

As one can probably tell from the example, that we have two key primitives across the three steps above: (1) ``similar-sheet'' and (2) ``similar-region'', which are crucial for us to correctly predict the target formula, and are the key technical challenges we address in this paper.

Specifically, we imagine spreadsheets and spreadsheet-regions as ``images'', where each cell is like a pixel. We then develop deep models to learn suitable ``dense vector representation'' for both spreadsheets and spreadsheet-regions, similar to how images can be represented as dense vectors~\cite{cnn, facenet, sphereface}. Using the vector-representations of spreadsheets and spreadsheet-regions, we can then quickly find both ``similar-sheet'' and ``similar-region'', leveraging standard approximate nearest neighbor (ANN) search for vectors~\cite{ann-1, ann-2}.


\begin{figure}[t!]
    \centering   \includegraphics[width=0.8\columnwidth]{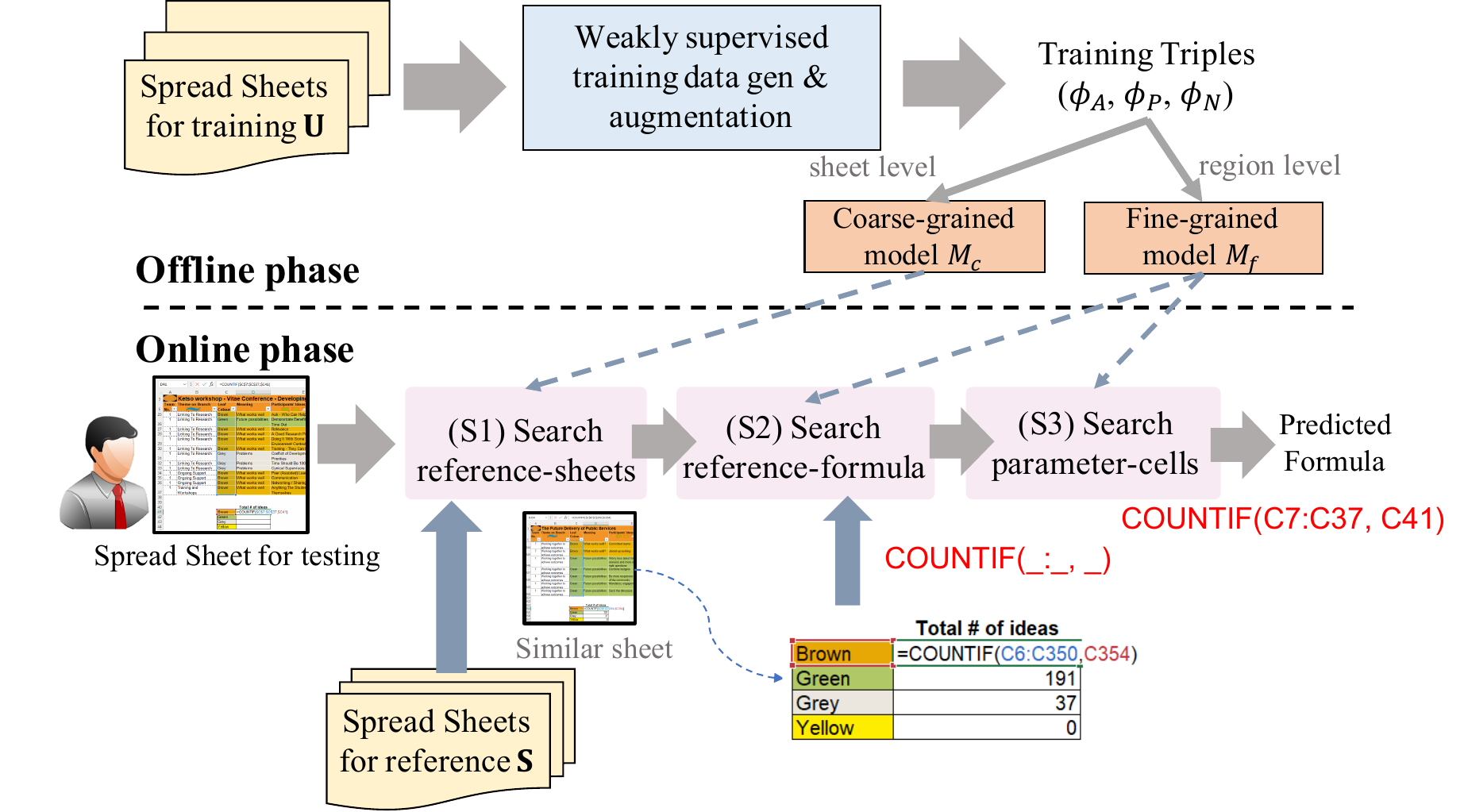}
    \caption{Overall system architecture of \sys.}
    \vspace{-3mm}
    \label{fig:overall}
\end{figure}

\stitle{System Architecture.} Figure~\ref{fig:overall} shows the overall architecture of our system, which contains (1) {offline steps}, shown in the top half of the figure, and (2) {online steps} that is shown in the lower half.

In the \underline{offline steps}, we first automatically generate training examples of similar vs. dis-similar spreadsheets, from a large crawl of 160K spreadsheets $\mathbf{U}$, using weak-supervision and based on a hypothesis-test we develop (Section~\ref{sec:traindatagen}). The training data $\mathbf{U}$ will then be used to learn spreadsheet representation models, for effective similar-sheet and similar-region search (Section~\ref{sec:representation}). Note that because our 160K spreadsheets are crawled from across the web, the representation-model is universal and applicable to all kinds of spreadsheets, where the learning step \emph{happens only once}. Given a new corpus of spreadsheets $\mathbf{S}$ in a new organization, we apply the representation models (learned on $\mathbf{U}$) to generate vector representations for each spreadsheet $S \in \mathbf{S}$, which we will then index offline.

The \underline{online phase} uses the three steps of (S1) search similar sheets, (S2) search reference formulas and (S3) search parameter-cells, which as explained in Example~\ref{ex:3steps} above, can accurately predict likely formulas in a target spreadsheet cell.



\subsection{Weakly-supervised training data generation} 
\label{sec:traindatagen}
In order to learn a spreadsheet-representation model, so that we can reliably predict ``similar-sheet'' and ``similar-region'', the first step is to generate positive and negative examples, that correspond to similar and dis-similar sheets/regions, respectively. Since manual labeling of examples is expensive and hard to scale, we propose to use \emph{weak-supervision} to automatically generate training data.





\begin{figure*}
    \centering   \includegraphics[width=\textwidth]{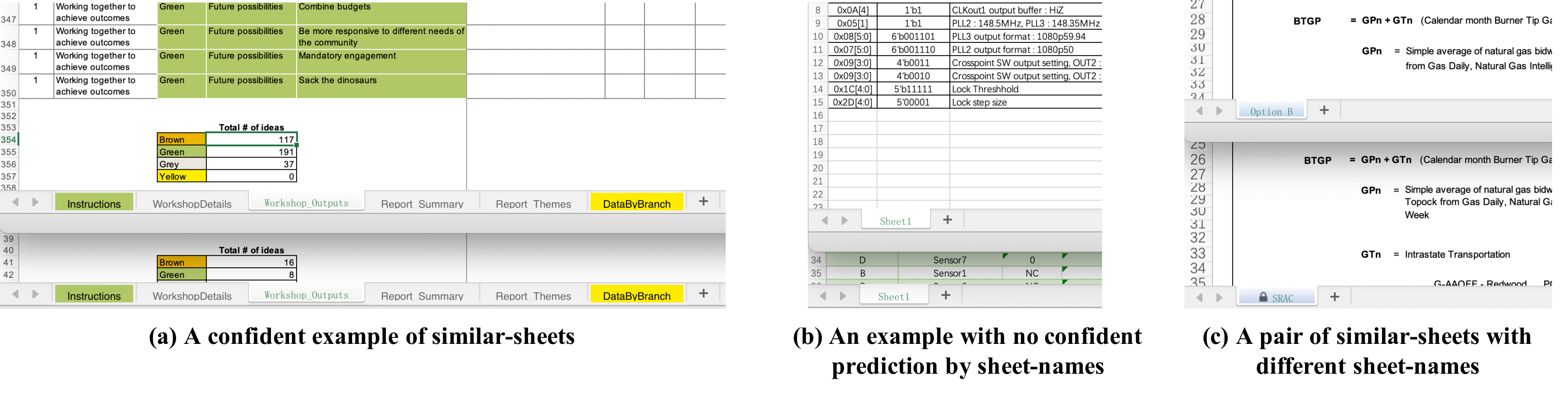}
    \vspace{-8mm}
    \caption{Weak-supervision using sheet-names. (a) Two files with two sequences of sheets, where all sheet-names are identical, which likely indicate that the sheets between the two files are similar-sheets. (b) Two files with only one sheet called ``Sheet1'' (a common name), for which weak-supervision is not confident that the two are similar-sheets. (c) Example similar-sheets that do not share the same sheet-name, which will be missed by our weak-supervision that performs hypothesis-tests on sheet-names.}
    \label{fig:weak-supervision}
\end{figure*}

\stitle{Find similar-sheets by sheet-names.} Each spreadsheet file, usually stored in the \textit{.xlsx} or \textit{.xls} format, 
typically contains multiple ``sheets''~\cite{worksheet}, where each sheet is a two-dimension grid, similar to a traditional table.
For example, each screenshot shown in Figure~\ref{fig:weak-supervision} is a spreadsheet file, where the tabs shown at the bottom of each file, such as ``\code{Instructions}'' and `\code{WorkshopDetails}'' in Figure~\ref{fig:weak-supervision}(a), are  sheets that belong to the same file.

Our key intuition for using weak-supervision to find similar-sheets, is that when two spreadsheet files, denoted as $F$ and $F'$, contain multiple sheets, denoted as $F = (s_1, s_2, \ldots s_n)$ and $F' = (s'_1, s'_2, \ldots s'_n)$, yet all the names of these sheets match exactly 1-to-1 with each other, or $name(s_i) = name(s'_i), \forall i \in [n]$, then these two files ($F$ and $F'$), as well as their corresponding sheets ($s_i$ and $s'_i$), are likely to be similar.

Figure~\ref{fig:weak-supervision}(a) shows such an example, where we have two separate spreadsheet files, stacked on top of each other. We can see that the two sequences of sheets from the two files have identical names, e.g.,  [``\code{Instructions}'',  `\code{WorkshopDetails}'', $\ldots$]. Intuitively, if the two spreadsheet files were unrelated, it is highly unlikely for them to share the identical sequence of sheet-names just by chance, so the two files are likely generated from the same source, following similar processes, such that each pair of sheets are similar-sheets.

However, on the flip side, because two spreadsheet files $F$ and $F'$ have sheets sharing identical names, does not mean that $F$ and $F'$ must be similar. Figure~\ref{fig:weak-supervision}(b) shows such a counter-example, where the two files both contain only one sheet called ``\code{Sheet1}'', which is a really common sheet-name (a default name generated by systems). That is not sufficient evidence for us to predict the two sheets to be related -- upon a closer inspection, the content in these sheets indeed looks very different.



\textbf{Weak supervision by hypothesis-tests.}
We clearly need to model the probabilities of seeing different sequences of sheet-names (common vs. rare names,  long vs. short sequences), for this to be reliable.
We therefore develop a weak supervision method, using a hypothesis test we propose. 

\textbf{}
Our null hypothesis $H_0$ states that two given spreadsheets $F$ and $F'$ are \emph{not similar} (a default position that should apply in the majority of cases), and for the sheet-names of $(s_1, s_2, \ldots s_n)$ and $(s'_1, s'_2, \ldots s'_n)$ to match is a just coincidence.  We then explicitly model the probabilistic process of drawing two sequences of sheets $(s_1, s_2, \ldots s_n)$ and $(s'_1, s'_2, \ldots s'_n)$ from a universe of sheet-names -- if we find the probability of two given sequences of names to collide is exceedingly small (e.g., smaller than typical $\alpha$ thresholds like 0.05), we reject $H_0$ and conclude that this is not a coincidence, the two files $F$ and $F'$, as well as their sheets, are likely similar. 

Specifically, let $p_i = Pr(name(s_i))$ be the probability of encountering the sheet-name $name(s_i)$, when we draw a random sheet from all sheets in the universe. Let $U$ be the universe of all spreadsheets, then $p_i = \frac{freq_U(name(s_i))}{|U|}$, where $freq_U(name(s_i))$ is the frequency of $name(s_i)$ in all spreadsheets in $U$, and $|U|$ is the total number of spreadsheets.

Given two sequences of sheets $(s_1, s_2, \ldots s_n)$ and $(s'_1, s'_2, \ldots s'_n)$, the chance of seeing the sheet-names in the second to match exactly with the first, can then be calculated as $\prod_{i \in [n]}{p_i}$ (because each time when we draw a sheet $s'_i$, the probability that its name matches the name from the first sequence, $name(s_i)$, is $p_i$, $\forall i \in [n]$).\footnote{We made a simplifying assumption of independence, which is analogous to 1-gram language-models to account for rare events (sheet-names). Considering joint distributions of sheet-names is an alternative to estimate probabilities here.}

Under the null-hypothesis, if we find the probability of our observation (two sequences of sheets sharing identical names), $\prod_{i \in [n]}{p_i}$, to be exceedingly small, e.g, smaller than the typical alpha of 0.05, we reject $H_0$ and conclude that  ($s_i$, $s_i'$) are similar sheets, for all $i \in [n]$.

We demonstrate this process of hypothesis-test more concretely, using the following example.

\begin{example} 
\label{ex:hypothesis-test}
For the example in Figure~\ref{fig:weak-supervision}(a), we find the first sheet-name ``\code{Instructions}'' to occur $100$ times from a universe of $100K$ sheets, so the probability of drawing a random sheet from the universe and seeing this name ``\code{Instructions}'', is $\frac{100}{100k}=0.1\%$. We do the same calculation for the second sheet-name in the example, ``\code{WorkshopDetails}'', which is rare that occurs $10$ times in total, so its probability is $\frac{10}{100K}$, etc. When we do the full cross-product, we get a final p-value of $\prod_{i \in [n]}{p_i}  = 10^{-13}$, which is smaller than $0.05$, making us to reject $H_0$ and conclude that the two sequences of sheets are similar.  

For the case in Figure~\ref{fig:weak-supervision}(b), because "\code{Sheet1}" is a common name that occurs $15k$ times out of a total of $100K$ spreadsheets, its p-value is $\frac{15k}{100K} = 0.15$, which is not sufficient for us to reject $H_0$, so we cannot conclude that the two sheets in Figure~\ref{fig:weak-supervision}(b) must be similar.
\end{example}

\textbf{Generate positive/negative training pairs for sheets and regions.} Using the hypothesis-test above as weak-supervision, we automatically generate positive and negative training pairs, for similar and dis-similar sheets as follows. 

\underline{Similar sheets.} To generate positive examples for similar sheets, we check pairs of spreadsheet files ($F, F'$), whose sequences of sheets $(s_1, s_2, \ldots s_n)$ and $(s'_1, s'_2, \ldots s'_n)$ have identical names. We mark all ($s_i$, $s_i'$) $\forall i\in[n]$ as positive examples (similar-sheets), if $\prod_{i \in [n]}{p_i} \leq \alpha$, (e.g., $\alpha = 0.05$), so that we can reject $H_0$, like discussed above.

To generate negative examples for similar sheets, it is sufficient to sample two random sheets as negative examples, because the chance of ``collision'' (the two being similar) in a large spreadsheet corpus is vanishingly small. To be extra safe, we add an even stronger requirement --  we sample pairs of spreadsheet files ($F, F'$), and only if the two sets of sheets $(s_1, s_2, \ldots s_n)$ and $(s'_1, s'_2, \ldots s'_n)$ do not share even one common sheet-name, do we then proceed to use pairs of $(s_i, s'_j)$ as negative examples (dis-similar sheets).



\underline{Similar regions.} Recall that in addition to similar-sheets, we also need to learn to detect similar-regions (used in our steps S2 and S3 in Section~\ref{sec:intuition}). 

To generate positive examples for similar-regions, we take a pair of similar-sheet ($s_i$, $s_i'$), and check for all formulas on $s_i$ and $s_i'$. If we have formula $f \in s_i$ and $f' \in s_i'$, where the locations of $f$ and $f'$ are identical, or $Loc(f) = Loc(f')$ (e.g., one is in \code{B59} of $s_i$ while the other is also in \code{B59} of $s_i'$), and furthermore their formula-expressions are also identical, or $f = f'$ (e.g., both are \code{SUM(A12:B40)}), then we are confident that the regions surrounding $Loc(f)$  and $Loc(f)$ must be similar-regions, which we will use as positive examples. 

To generate negative examples for similar-regions, we simply take the positive examples ($Loc(f), Loc(f')$) above, and shift one of the locations $Loc(f')$ until it hits a different formula $g$ with $g \neq f$, which is when we stop and mark ($Loc(f), Loc(g)$) as a negative example for similar-regions.





\textbf{Discussion.} 
We find our hypothesis-test based weak-supervision to be of high accuracy -- e.g., when we manually sample and verify positive/negative examples so generated, we find the precision of positive/negative labels to be over 0.95. 

However, at the same time, because this is a weak-supervision method, this is meant to only catch ``confident'' positive/negative examples, but it will miss out on many others (i.e., high precision but low recall, as we will show in our experiments in Section~\ref{sec:exp}). For example, Figure~\ref{fig:weak-supervision}(c) shows two files, where their sheet-names are different, yet their actual content is similar (thus a positive example for similar-sheets). Weak supervision will not be able  to catch such cases as it relies solely on strong sheet-name overlap, and does not consider actual content -- using spreadsheet content to detect similar-sheet and similar-region, for both high precision and recall, is the goal of our learned-representation models below.

\begin{figure*}
    \centering   \includegraphics[width=1.03\textwidth]{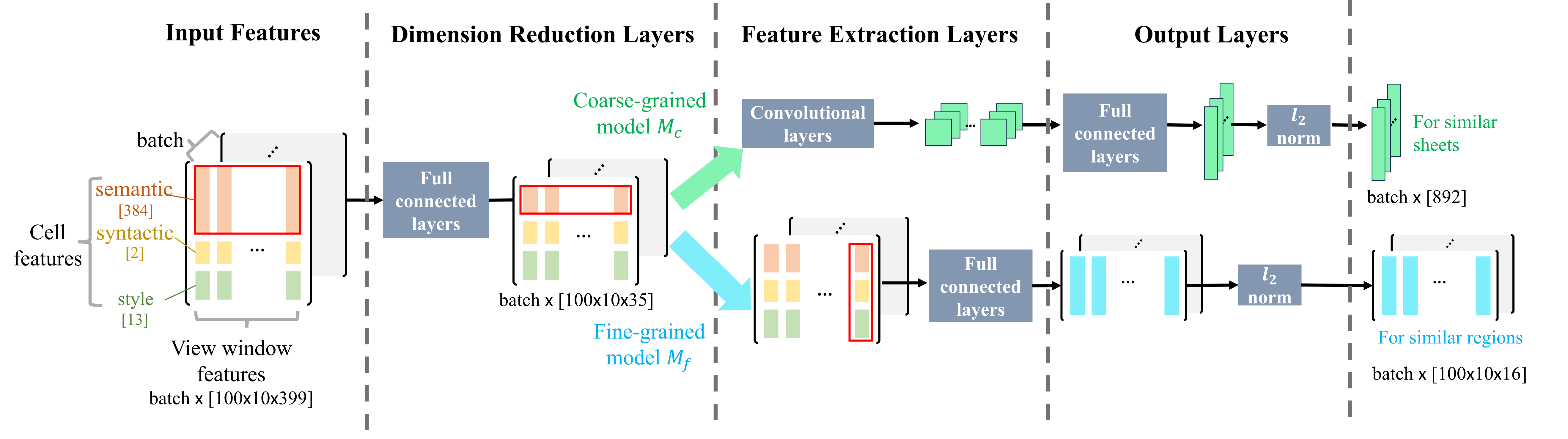}
    \caption{Our model architecture for spreadsheet representations. The model branches into two towards the end, which are (1) a coarse-grained model $M_c$, for similar-sheet detection; and (2) a fine-grained model $M_f$, for similar-region detection.}
    \label{fig:arc}
\end{figure*}

\subsection{Training data augmentation}
\label{sec:data-augmentation}

In order to improve model generalizability tables of varying sizes, we use data augmentation~\cite{da-1, da-2} to enhance our training dataset. Specifically, for a pair of similar sheets or regions (positive examples), we randomly remove some fraction of rows and columns from one sheet/region in the pair, and continue to use the resulting pair as positive examples, where the idea is that two sheets/regions sharing the same template will continue to be considered ``similar'', even if a small number of rows and columns are inserted/removed (common in spreadsheets as users frequently insert/remove rows and columns in an existing spreadsheet for their needs).



For similar sheets, we augment all positive pairs, by removing rows and columns with equal probability $p$ ($p$ is randomized between $0-10\%$ for each sheet). 
For similar regions, we find it beneficial to augment more carefully, e.g., by removing only the bottom-most rows and right-most columns using the same probability $p$, which tends to remove data rows and columns, while keeping the table structures (e.g., column headers and entity columns) intact. 
For similar-sheets, we only augment a random subset of all regions ($20\%$).

\subsection{Models for spreadsheet representation}   \label{sec:representation}

Now that we have harvested lots of positive and negative examples for similar-sheet and similar-regions, we proceed to train our spreadsheet-representation models. The goal here is to represent sheets and regions as dense vectors using the models, so that when given a new pair of sheets $(s, s')$ or regions $(Loc(c), Loc(c'))$, we can quickly and accurately compute their similarity.

Given the structure of spreadsheets where cells are laid out in a two-dimensional grid, it is natural to think of them as ``images'' where pixels are also organized in a two-dimensional manner. And for images there is a large literature from computer-vision for image representations~\cite{facenet, cosface, sphereface}. 

Our representation models for spreadsheets are inspired by classical architectures in computer-vision, but are tailored specifically to spreadsheets. Figure~\ref{fig:arc} shows the main architecture of our models, which contains input, dimension reduction, feature extraction, and output layers, respectively.  

Recall that we need to have two models for ``similar-sheet'' and ``similar-region'', respectively, which is why our architecture branches out in the feature-extraction layers, into (1) coarse-grained models (for similar-sheet) and (2) fine-grained models (for similar-region).  

Next, we will go over these layers in turn below.


\subsubsection{Input features.} \label{sec:feature} \hfill\\
Our first step in generating representations for spreadsheets, is to represent spreadsheets and spreadsheet-regions as input feature vectors, so that we can feed them into deep models for representation learning. Because spreadsheets and spreadsheet-regions are two-dimensional grids consisting of cells, we will first describe how we generate input features for each cell.


\stitle{Cell Features.} Unlike database tables where each cell contains just a value, spreadsheet cells contain not only content but also rich style features (color, font, font-size, borders, etc.) for human consumption, which are all salient features for spreadsheet-representations and for finding similar-sheets (e.g., in Figure~\ref{fig:formula-ex}). 
Therefore, we design ``\emph{content-features}'' and ``\emph{style-features}'' for each spreadsheet cell, described below.

\etitle{Content features $\gamma^{c}$.} We use content features to represent the content in a cell, which in turn have \emph{semantic} and \emph{syntactic} features.


\begin{itemize}[noitemsep,topsep=0pt,leftmargin=*]
    \item \textit{Semantic features.} We directly apply a pre-trained natural-language embedding Sentence-BERT\cite{sentencebert}, to produce a vector representation of each cell value, so that for example, ``\code{USA}'' and ``\code{Canada}'' will be close by in this feature space. (This can also be easily replaced with alternative embedding methods such as Glove~\cite{glove}, as we show in our experiments)

    \item \textit{Syntactic features.} We represent the data type of each cell (numeric, text, empty, etc.), as a categorical feature. In addition, we represent the syntactic-patterns of values (e.g., ``\code{DDDD-DD-DD}'' for ``\code{2020-01-01}'') also as features.
\end{itemize} 

\etitle{Style features $\gamma^{s}$}. We represent rich styles encoded in spreadsheet cells, including background color, font color, font sytle (bold/italic/etc.), font size, cell size (height and width) as a feature vector $\gamma^{s}$.

    

Our final feature vector for each cell $C$, denoted by $\gamma(C)$, is then the simple concatenation of $\gamma^{c}(C)$ and $\gamma^{s}(C)$.\footnote{We note that formulas within cells can also be useful cell-level features. However, because we study spreadsheets as static artifacts, where we do not know the order in which formulas are created, we choose not to use such features in order to avoid using formula features that are not present when users author a test formula, so as to not over-estimate the effectiveness of our method.}

\stitle{View window.} 
Now that we have a feature vector $\gamma(C)$ for each cell, we need to represent spreadsheets and spreadsheet-regions. 

\begin{figure}[t!]
    \centering   \includegraphics[width=0.8\columnwidth]{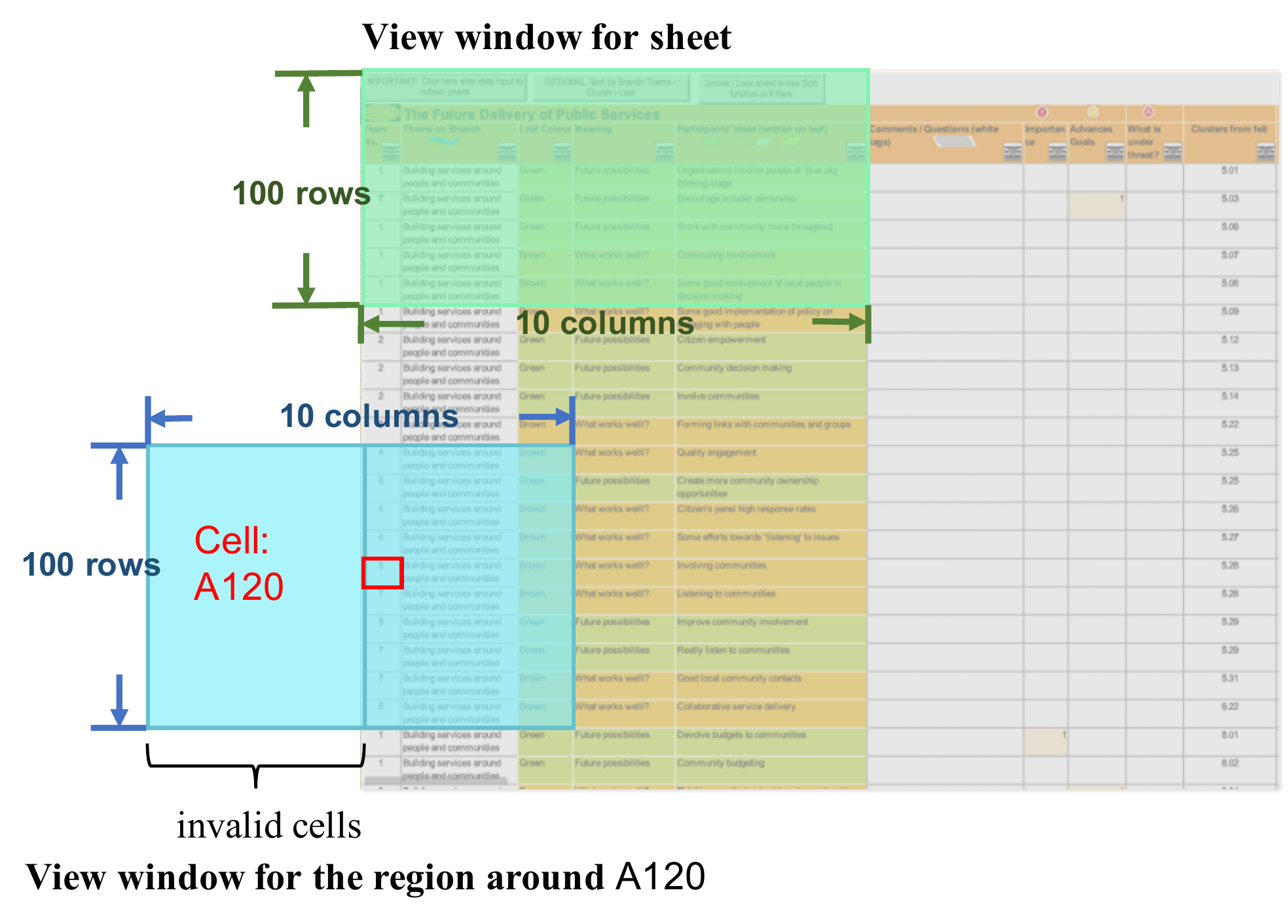}
    \caption{Example of view windows. The blue box at the bottom represents a region centered at \code{A120}. The green-box uses the view-window at top-left to represent the entire sheet.}
    \label{fig:viewwindow}
\end{figure}

Because there are no explicit table boundaries in spreadsheets like we mentioned earlier (e.g., there may be multiple irregularly-shaped tables on the sheet), we use a fixed window of $n_r$ rows and $n_c$ columns as our ``view window'' (similar to a view window that human eyes can focus on), e.g., with $n_r = 100$ rows and $n_c = 10$ columns. This view window can move around in a spreadsheet to represent different regions,  like shown in Figure~\ref{fig:viewwindow}.


To represent the surrounding region of a cell $C$, we use the $(n_r \times n_c)$ view window, centered at the location of $C$, denoted as $V(C)$. The blue-box in Figure~\ref{fig:viewwindow} shows such an example -- to represent the region around \code{A120}, we use the $(n_r \times n_c)$ blue-box centered around  \code{A120}, as our view window $V(\code{A120})$. 

To represent an entire spreadsheet, we simply use a $n_r \times n_c$ view window starting from the top-left corner of a spreadsheet, like shown by the green box in Figure~\ref{fig:viewwindow}, which is taken as a representative region of the entire sheet.

\stitle{Input vector.} Finally, the input vector $\Phi_{V}$ for a view window $V$, representing a spreadsheet or spreadsheet-region, is simply the cell-features for all cells in the view window stacked in 2D manner, denoted as $\{\gamma(C) | \forall C \in V \}$. We maintain the same 2D structure for a 2D table region, with each cell represented as a vector depicted in the vertical direction, like shown in the left-most segment of Figure~\ref{fig:arc}.




\subsubsection{Dimension reduction layers.}\hfill\\
We now move to the dimension-reduction layers, which is shown in the second segment of Figure~\ref{fig:arc}. 

These dimension-reduction  layers are needed, because our input features of $\gamma(C)$ for each cell $C$, is a concatenation of different types of features, resulting in hundreds of dimensions (Sentence-BERT alone has hundreds of dimensions).  We use the dimension-reduction  layers (with shared weights for each cell) to distill the features important for our task of spreadsheet representation. These are implemented as Multi-Layer Perceptron (MLP) layers.


\subsubsection{Feature extraction layers.}\hfill\\
At the feature-extraction layers, shown in the third segment of Figure~\ref{fig:arc}, we now branch out to two variants of the model, which we call ``\emph{coarse-grained}'' and ``\emph{fine-grained}'' models, that are specialized for ``similar-sheet'' and ``similar-region'', respectively. In order to see why such a specialization is needed, we revisit our running example below.

\begin{example}
\label{ex:fine-grained-vs-coarse-grained}
    Consider our example in Figure~\ref{fig:formula-ex}. To predict a formula in \code{D41}, we need to find the similar-sheet for the sheet in Figure~\ref{fig:formula-ex}(a) on the left, which is the sheet in Figure~\ref{fig:formula-ex}(b) on the right. We also need to find similar-regions, which for the region centered at \code{D41} on the left, represented by the view-window $V_l(\code{D41})$, is the region centered at \code{D354} on the right, denoted by $V_r(\code{D354})$.

    However, as we can intuitively see, for similar-sheet (e.g., Figure~\ref{fig:formula-ex}(a) and Figure~\ref{fig:formula-ex}(b)), the comparison will likely need to be ``fuzzy'', because while a pair of similar-sheets may have a lot in common, they may still have different number of columns and rows (e.g., 37 rows vs. 350 rows in this example), because the two sheets are often populated with different content.

    In contrast, for similar-region comparisons (e.g., $V_l(\code{D41})$ and $V_r(\code{D354})$), our comparison will need to be ``precise'', because if we were to  shift slightly, and use the region $V_l(\code{D\color{black}\textbf{{355}}})$ that shifts one cell to the down from $V_l(\code{D354})$, in terms of ``fuzzy'' similarity the two regions are still very similar, but this slight shift will lead to a different and incorrect formula (e.g., we will recommend the formula for ``\code{Green}'' in \code{D355} from the right, instead of the desired formula for ``\code{Brown}'' in \code{D354}).  
    
\end{example}


This example motivates us to use the same feature representations from previous layers, and then branch out to two specialized models, for (1) coarse-grained spreadsheet-representation, where ``cell-by-cell alignment'' is not important, for ``fuzzy'' sheet-level similar-sheet search; and (2) fine-grained region representation, where cell-by-cell alignment is crucial, for ``precise'' region-level similar-region search.

Architecture-wise, for the coarse-grained models, we use the classical Convolutional Neural Network~\cite{cnn}, which is translation invariant, and insensitive to shifts of rows and columns, but at the same time it also ``blurs'' the boundary between cells that creates a ``fuzzy'' representation of spreadsheets (e.g., even though the color schemes of Figure~\ref{fig:formula-ex}(a) and Figure~\ref{fig:formula-ex}(b) are not identical cell-by-cell, they are similar enough in a global and fuzzy sense, which is what the coarse-grained model can capture).

The architecture for the fine-grained models, on the other hand, will suffer if CNN was used, because a blurry convoluted picture would make it hard for us to find the precise locations of formulas and parameter-cells. As such, we use fully-connected networks here, which preserve cell boundaries, so that we can identify precise similar-regions  (e.g., $V_l(\code{D41})$ and $V_r(\code{D354})$ as discussed in Example~\ref{ex:fine-grained-vs-coarse-grained}).

    

\subsubsection{Output layers.}\hfill\\
The output from the feature extraction layers above, are now dense vectors representing a coarse-grained and fine-grained views of a spreadsheet region. 
We further perform $L_2$ normalization here, so that the resulting dense vectors can be directly used to compute similarity, between two spreadsheets or spreadsheet-regions. 

\subsection{Train spreadsheets as ``face recognition''} \label{sec:opt}
Using the model architectures above, we can now proceed to train spreadsheet representations, so that we can effectively detect similar vs. dis-similar spreadsheet regions. 

Here we take inspiration from classical computer-vision techniques for ``face recognition'' (e.g.~\cite{facenet, sphereface, cosface}), where the key problem is to find ``similar faces'' belonging to the same person, from an ocean of faces in a database (typically with the help of manually labeled similar vs. dis-similar faces), where the differences between faces can be subtle. Our problem of ``similar-sheet'' and ``similar-region'' is similar in spirit, in that we need to teach the models to learn to detect subtle differences between spreadsheets.


\etitle{Semi-hard triplet learning for similar-spreadsheets.} 
In the face-recognition literature,  the \emph{semi-hard triplet-learning} approach pioneered by FaceNet~\cite{facenet}, is particularly effective. Intuitively, in semi-hard learning, in each step, the algorithm selects many triplet examples $(A, P, N)$, where $A$ is an anchor (a reference face), $P$ is a positive example to the anchor (a similar face), and $N$ is a ``semi-hard'' negative example (a different face, that nevertheless looks somewhat similar to $A$). By carefully selecting $(A, P, N)$ triples as training progresses, one can make sure that the negative example $N$ is always hard and useful to learn from, but at the same time it is not too hard for the gradient to completely break down (or too easy for the model to not learn much). 

We adapt the triplet learning framework to the spreadsheet domain. Specifically, we use the triplet-loss as our training objective for our representation model shown in Figure~\ref{fig:arc}:
\begin{equation}
l_{triplet} = max(\Vert\phi_{A} - \phi_{P}\Vert^2  - \Vert\phi_{A} - \phi_{N}\Vert^2 + m, 0)  
\label{eqn:loss}
\end{equation}
where $(A, P, N)$ are three spreadsheet-regions, with $(A, P)$ being a pair of automatically generated positive examples from our weak-supervision step (Section~\ref{sec:traindatagen}), and $(A, N)$ a pair of negative examples from the same process.


In order to compute the loss in Equation~\eqref{eqn:loss}, we feed the input features representing $A, P, N$, respectively, into the model of Figure~\ref{fig:arc}, and then use the dense vectors produced from the output layers of the model, denoted by $\phi_{A}, \phi_{P}, \phi_{N}$, respectively, to compute the triplet loss.

Here, $\Vert\phi_{A} - \phi_{P}\Vert^2$ measures the $L_{2}$ distance between the representations for the two positive examples $\phi_{A}$ and $\phi_{P}$, which after training should ideally be small, while $\Vert\phi_{A} - \phi_{N}\Vert^2$ is the $L_{2}$ distance between two negative examples, which should ideally be large.  The extra $m$ term in Equation~\eqref{eqn:loss} refers to \textit{margin}, which is a hyper-parameter that encourages the model to continue to push negative examples $N$ further away from positive examples $P$, relative to anchor $A$, until there is a safe margin $m$ to differentiate the two. 


Like training models for face-recognition, during each step in training, we continuously select semi-hard triples $(A, P, N)$ that are the most informative for models to learn, based on the current state of the models.  Specifically, we sample triples $(A, P, N)$ whose loss satisfies $0 < l_{triplet}(A, P, N) < m$, which ensures that the selected triple $(A, P, N)$ is neither too hard (with loss greater than $m$), nor too easy (with loss equals 0).

This training process is summarized in Algorithm~\ref{alg:offline}. After the training process converges,  we produce a coarse-grained representation model for similar-sheets, denoted by $M_c$, (shown at the top of Figure~\ref{fig:arc}), and a fine-grained representation model for similar-regions, denoted by $M_f$ (shown in the bottom half of Figure~\ref{fig:arc}).



\iftoggle{fullversion}
{
    \begin{figure*}
        \centering   \includegraphics[width=1\textwidth]{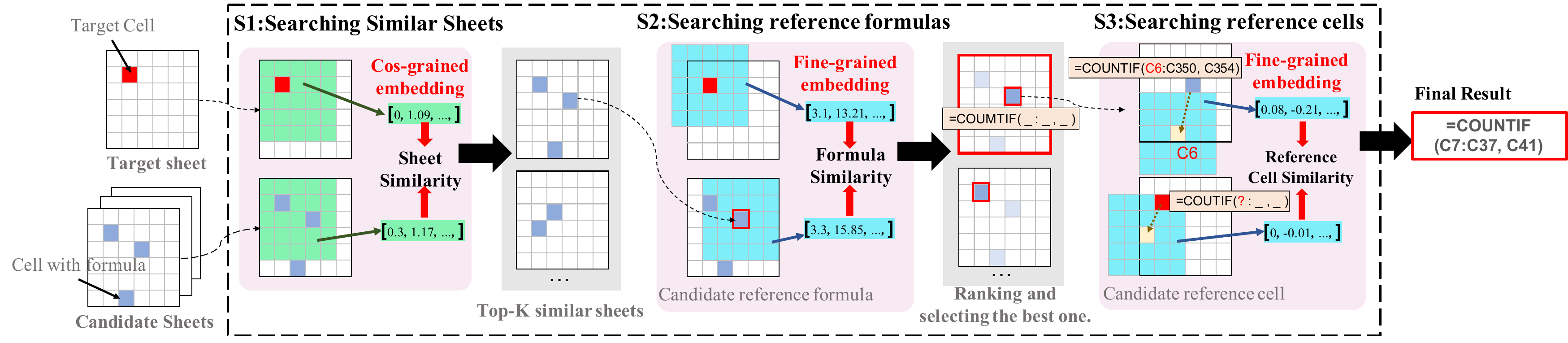}
        \caption{The online formula prediction process of \sys. }
        \label{fig:framework}
    \end{figure*}
}
{

}

\input{algs/offline}

\iftoggle{fullversion}
{

\input{algs/end2end}

The full pseudo-code for the steps in the offline phase, can be found in Algorithm~\ref{alg:end2end}.
}
{
The full pseudo-code for the steps in the offline phase, can be found in an extended version of the paper~\cite{full}.
}

\subsection{\sys: Putting it together} \label{sec:online}
Now that we have described each component, we will walk through our system end-to-end, explaining how the components connect with each other. For this purpose, we will revisit our architecture diagram in Figure~\ref{fig:overall}.

In the offline training time, we crawl a large corpus of 160K spreadsheets, denoted by $\mathbf{U}$, from across the web. We develop a weak-supervision method based on hypothesis tests, which when applied to the spreadsheet corpus,  can automatically generate large amounts of positive and negative examples for similar-sheets/similar-regions (Section~\ref{sec:traindatagen}). The positive/negative examples so produced are fed into our representation-learning models (Section~\ref{sec:representation}), which uses semi-hard triplet learning to train two related models, one coarse-grained model $M_c$, for similar-sheet detection, and one fine-grained model $M_f$ for similar-region detection, respectively, like shown by the two orange boxes in Figure~\ref{fig:overall} (Section~\ref{sec:opt}). 

Because the models so produced are universally applicable to spreadsheets across the web $\mathbf{U}$, the training process only happens once. For a new collection of spreadsheets of interest (e.g., from within an organization), denoted as $\mathbf{S}$, we only need to perform inference on each spreadsheet $S \in \mathbf{S}$. Specifically, (1) we generate a dense vector representation $M_c(S)$ for the entire sheet $S$, which is a sheet-level ``signature'' analogous to LSH. We add each $M_c(S)$ into a standard ANN index, denoted as $Idx_c(\mathbf{S}) = \{M_c(S) | S \in \mathbf{S} \}$, for efficient retrieval at online time. Furthermore, (2) we can optionally also index promising regions $V(C)$ from existing spreadsheets, whose center-cell $C$ contain formulas that may be used as reference-formulas. We perform inference using the fine-grained model $M_f$ on such region $V(C)$, and add the resulting vector $M_f(V(C))$ to a region-level fine-grained index $Idx_f$, defined as: $Idx_f(\mathbf{S}) = \{M_f(V(C)) |  S \in \mathbf{S}, C \in S,  \text{C has a formula} \}$.

\iftoggle{fullversion}{
    At online time, we perform the three steps outlined in Section~\ref{sec:intuition} (also shown in Figure~\ref{fig:framework}), where the coarse-grained index $Idx_c(\mathbf{S})$ can be used to quickly find similar-sheets (step S1), and the fine-grained index $Idx_f(\mathbf{S})$ can be used to quickly find similar-regions for formulate template (step S2), and then used again to find parameter-cells that can be filled into formulate templates (step S3). These ANN indexes $Idx_c(\mathbf{S})$ and $Idx_f(\mathbf{S})$, are the key reasons for us to find similar-sheets and regions, both accurately and efficiently.

    We note that in this study, we focus exclusively on model-based predictions of formulas. There are additional opportunities to further improve prediction quality, such as  post-processing predicted formulas, by validating formulas against the local context of the spreadsheet. These are optimizations orthogonal to our learning-based approach, which we will leave as future work.
      
}{
    At online time, we perform the three steps outlined in Section~\ref{sec:intuition} (also shown in the lower part of Figure~\ref{fig:overall}), where the coarse-grained index $Idx_c(\mathbf{S})$ can be used to quickly find similar-sheets (step S1), and the fine-grained index $Idx_f(\mathbf{S})$ can be used to quickly find similar-regions for formulate template (step S2), and then used again to find parameter-cells that can be filled into formulate templates (step S3). These ANN indexes $Idx_c(\mathbf{S})$ and $Idx_f(\mathbf{S})$, are the key reasons for us to find similar-sheets and regions, both accurately and efficiently. 

    We note that in this study, we focus exclusively on model-based predictions of formulas. There are additional opportunities to further improve prediction quality, such as  post-processing predicted formulas, by validating formulas against the local context of the spreadsheet. These are optimizations orthogonal to our learning-based approach, which we will leave as future work.
    }

%% file: algs/offline.tex
\begin{algorithm}[t!]
\small
    \caption{Offline training}
    \label{alg:offline}
    \KwIn{A large corpus of spreadsheets $\mathbf{S}$}
    \KwOut{Coase-grained model $M_{c}$, Fine-grained model $M_{f}$}
    
    Initialize $M_{c}$, $M_{f}$ with random parameters.

    $Pr_{s_{P}}, Pr_{s_{N}}, Pr_{r_{P}}, Pr_{r_{N}} = TrainDataGen(\mathbf{S})$

    \For{each $episode \in [1, T]$}{
        $(\phi_{c_{A}},\phi_{c_{P}},\phi_{c_{N}})=SHSample(Pair_{s_{P}},Pair_{s_{N},M_{c}})$

        $(\phi_{f_{A}},\phi_{f_{P}},\phi_{f_{N}})=SHSample(Pair_{r_{P}},Pair_{r_{N},M_{f}})$

        $l_{triplet_{c}} =  max(\Vert\phi_{c_{A}} - \phi_{c_{P}}\Vert^2  - \Vert\phi_{c_{A}} - \phi_{c_{N}}\Vert^2 + m, 0)$

        $l_{triplet_{f}} =  max(\Vert\phi_{f_{A}} - \phi_{f_{P}}\Vert^2  - \Vert\phi_{f_{A}} - \phi_{f_{N}}\Vert^2 + m, 0)$

        Perform SGD training on $M_{c}$ with $l_{Triplet_{c}}$.

        Perform SGD training on $M_{f}$ with $l_{Triplet_{f}}$.
    }
\end{algorithm}

	
        
        
 
        
            
            
            
            
    

%% file: algs/end2end.tex
\begin{small}

\begin{algorithm}[t!]
	\caption{\sys: End-to-end} 
        \label{alg:end2end}
        \KwIn{Target sheet $S_{T}$, Target cell $C_{T}$, Candidate sheets $\mathbb{C}$ and Training workbooks $W_{train}$.}
        \KwOut{formula $\overline{F}(R)$}
        
        Initialize number of candidate similar sheets $K$.
        
        Initialize size of the neighborhood $d$.

        \BlankLine
        \tcp{offline phase}
        $M_{c}, M_{f} = OfflineTraining(W_{train})$

        \BlankLine
        \tcp{S1:searching similar sheets}
        \Begin{ 
        $\phi(S_{T}) = M_{c}(S_{T})$
        
        \For{each $S_{\mathbb{C}_{i}} \in \mathbb{C}$}{
            $\phi(S_{\mathbb{C}_{i}}) = M_{c}(S_{\mathbb{C}_{i}})$
        }
        $\mathbb{S} = Faiss(\phi(S_{T}), \{\phi(S_{\mathbb{C}_{i}})\}, K)$}
        \BlankLine
        \tcp{S2:searching reference formulas}
        \Begin{ 
            $\phi(C_{T}) = M_{f}(C_{T}, S_{T})$
            
            \For{each $S_{i} \in \mathbb{S}$}{
                \For{each formula $F_{S_{i}, j}$ on $S_{i}$}{
                    $\phi(C_{F_{S_{i}, j}}) = M_{f}(C_{F_{S_{i}, j}}, S_{i})$
                }
            }
                
            $\overline{F_{ref}}(R_{ref}) = Faiss(\phi(C_{T}, \{\phi(C_{F_{S_{i}, j}})\}, 1, \theta)$
        }

        \BlankLine
        \tcp{S3:searching reference cells}
        \Begin{
            Initialized reference cells $R_{T}$ as empty
            
            \For{each $C_{r_{i}} \in R_{ref}$}{
                $\phi(C_{r_{i}}) = M_{f}(C_{r_{i}})$
            
                $Row_{T_{i}} = Row(C_{r_{i}}) - Row(C_{F_ref}) + Row(C_{T})$ 
            
                $Col_{T_{i}} = Col(C_{r_{i}}) - Col(C_{F_ref}) + Col(C_{T})$
                
                \For{each $r \in Range(Row_{T_{i}} - d, Row_{T_{i}} + d)$}{
                
                    \For{each $c \in Range(Col_{T_{i}} - d, Col_{T_{i}} + d)$}{
                        $C_{cand_{r,c}}$ is a cell in $S_{T}$ at $(r, c)$.
                        
                        $\phi(C_{cand_{r,c}}) = M_{f}(C_{cand_{r,c}})$
                    }
                }
            $R_{T_{i}} = Faiss(\phi(C_{r_{i}}), \{\phi(C_{cand_{r,c}})\}, 1)$
            
            $R_{T}.add(R_{T_{i}})$
            }
        }

        $\overline{F}(R) = \overline{F_{ref}}(R_{T})$
\end{algorithm}

\end{small}

%% file: experiment.tex
\section{Experiments}
\label{sec:exp}
We conduct extensive evaluation 
on real spreadsheet data to evaluate the efficiency and effectiveness of different approaches. Our benchmark data is available at~\cite{data_repo} to facilitate future research.

\subsection{Experimental setup}
\label{sec:exp-setup}

\stitle{Datasets.} 
For this study, we crawled 160K spreadsheets (``\code{.xlsx}'' files) from the public web, denoted by $\mathbf{U}$, to train our representation models $M_c$ and $M_f$.
We hold out spreadsheets from a few
large fortune-500 organizations (described in more detail below), denoted by $\mathbf{T}$, to test formula-recommendation. The test spreadsheet corpora $\mathbf{T}$ are held \emph{completely separate} from the training corpora $\mathbf{U}$, and are therefore \emph{not} seen when training $M_c$ and $M_f$, so that we can test model generalizability to new and unseen spreadsheets.

Our holdout test spreadsheets $\mathbf{T}$ come from the following four domains:



\etitle{Cisco}. Cisco is a large technology company. We use spreadsheets crawled from the public-facing ``\code{cisco.com}'' domain as test data in our experiments. 

\etitle{PGE}. PGE is a large energy company. We similarly use spreadsheets from the public-facing   ``\code{pge.com}'' domain as our test data.

\etitle{TI}. Texas Instruments (TI) is a semiconductor company, and we use data from the  ``\code{ti.com}'' domain. 

\etitle{Enron}. The Enron Corpus~\footnote{https://github.com/SheetJS/enron\_xls} is a large spreadsheet corpus extracted from the Enron Corporation. This spreadsheet corpus has been used in a number of prior studies~\cite{enron-1, enron-2}. 

Let $\mathbf{T_d}$ be the spreadsheet corpora from an enterprise domain $d$ above.  Recall that in our approach outlined in Figure~\ref{fig:overall},  to predict formula in a new spreadsheet $S \in \mathbf{T_d}$, our models $M_c$ and $M_f$ (trained on the separate $\mathbf{U}$) will not need to be retrained, and we only need to perform inference calls using $M_c$ and $M_f$ on spreadsheets that already exist in the same enterprise, referred to as the ``reference set'' $\mathbf{S_d} \subset \mathbf{T_d}$,
to identify similar sheets for recommendations.

For each enterprise corpora $\mathbf{T_d}$ above, we select a subset of spreadsheets and sample formulas from these spreadsheets as our test cases for formula predictions\footnote{For each test spreadsheet, we sample at most $10$ formulas to avoid over-representation, as some spreadsheets can have large (thousands) of formulas.}, and using the remaining spreadsheets in the same $\mathbf{T_d}$ as our reference set $\mathbf{S_d}$. 
We test two ways of selecting test spreadsheets:

\etitle{(1) Random}: we randomly sample 10\% of spreadsheets from each $\mathbf{T_d}$ as tests and use the remaining spreadsheets as reference $\mathbf{S_d}$;

\etitle{(2) Timestamp}: we order all spreadsheets in $\mathbf{T_d}$ by last-modified timestamps, and select the 10\% of spreadsheets most recently edited as our tests, with the remaining as reference $\mathbf{S_d}$.

Note that ``timestamp'' setting is more challenging but also realistic, as it models the real usage scenario where for a newly edited spreadsheet, we would hope to rely on previously created spreadsheets as ``references'', to recommend formulas on the new spreadsheet. We find our approach to be effective in both  ``random'' and ``timestamp''. In the remainder of the paper, we report results in the ``timestamp'' setting by default, and will defer results in the ``random'' setting to a full version of the paper. 

Table \ref{tab:datainfo} summarizes the key statistics of our test dataset across all four enterprise domains.

\input{tables/tableinfo}

\stitle{Evaluation Metrics.} We evaluate both the quality and efficiency of different methods for formula recommendation.

\etitle{Quality.} For each test case, if an algorithm $A$ predicts a formula, we compare it with the ground truth and mark it as a "hit" when the two match \emph{exactly}. 

In a test set with $n$ test cases, we count the number of cases for which an algorithm $A$ produces a prediction, denoted as $n_{pred}$ (note that $A$ may not make a prediction in all cases, if it is not confident). We count the number of predicted cases that are ``hits'' (exact match with the ground-truth), denoted as $n_{hit}$. The quality of the predictions can then be evaluated using the usual precision/recall/F1 measures:
$$recall = \frac{n_{hit}}{n}, \qquad precision = \frac{n_{hit}}{n_{pred}}$$
$$F1 = \frac{2*recall*precision}{recall+precision}$$


\etitle{Efficiency.} We measure the latency of algorithms using wall-clock time. All experiments were conducted on a Ubuntu 18.04 Linux VM with 24 vCPU cores and 188G memory. 

\stitle{Methods Compared}. We compare the following methods.

\etitle{Mondrian}~\cite{mondrian}. Mondrian is a novel method to detect the layout of spreadsheets by  clustering similar spreadsheets. A graph-node matching algorithm is proposed, which uses a hand-crafted similarity function to detect similar sheets. While Mondrian is not designed for formula-prediction, its graph-matching-based sheet-clustering is comparable to our learned approach to similar-sheets. 
We use authors original implementation on GitHub~\cite{Mondrian_github} for this comparison. 

\etitle{SpreadsheetCoder~\cite{spreadsheetcoder}}. While we were unable to run the code from  SpreadsheetCoder~\cite{spreadsheetcoder} (their GitHub repo explicitly mentions that their code is not meant to be runnable\footnote{\url{https://github.com/google-research/google-research/tree/master/spreadsheet_coder}}), we found that this technique  is already implemented in Google Sheets~\cite{gsbs}, as a formula recommendation feature that predicts formulas based on natural-language contexts. We therefore randomly sampled 180 formulas, and manually invoke these test cases in Google Sheets to perform this comparison. 


\etitle{GPT~\cite{GPT}}. Language models like GPT are capable of understanding tables~\cite{li2023table, zhang2024tablellm, lu2024large} and perform table tasks. To predict formulas using GPT, we use prompt-engineering techniques, such as Chain-of-thought reasoning (COT)~\cite{cot}, and Retrieval-Augmented-Generation (RAG)~\cite{rag} that embeds and retrieves similar sheets/regions to dynamically select best few-shot examples.

Specifically, we vary our GPT prompts along 4 dimensions:
\begin{itemize}[noitemsep, leftmargin=*]
\item[] (1) \textbf{Example Selection} (3x settings):  
    \begin{itemize}
      \item \underline{Zero-shot}:  we use no demo examples in this setting;
      \item \underline{Few-shot, using-common-formula}:  we use few-shot examples that consist of formulas commonly found in spreadsheets  (e.g., SUM and AVG) in this setting;
      \item \underline{Few-shot, using-RAG-formulas}:  we use few-shot examples dynamically retrieved from similar spreadsheets (using Glove embedding to represent spreadsheets and regions, indexed using ANN techniques FAISS), following a Retrieve-Augmented-Generation (RAG) paradigm~\cite{rag}.
    \end{itemize}
\item[] \add{(2)} \textbf{\add{Chain of Through}} \add{(2x settings)}: 
    \begin{itemize}
      \item  \underline{\add{With COT}}:  \add{we ask the model to decompose the task, and reason step-by-step in this setting, following a Chain-of-Thought (COT) approach proposed in~\cite{cot}}
      \item  \underline{\add{Without COT}}:  \add{we ask the model to provide answers directly, without using COT.}
    \end{itemize}
\item[] \add{(3)} \textbf{\add{Table regions}} \add{(2x settings): }
    \begin{itemize}
      \item  \underline{\add{Precise-table-region}}: \add{ we provide all cells within the target table boundary, as table context in the prompt, which allows the model to focus on the table of interest.}
      \item  \underline{\add{Large-sheet-region}}: \add{ we provide all cells within a large N by M region (which may include more than one table) as table context in the prompt, to allow the model to infer cross-table fomulas.}
    \end{itemize}
\item[] \add{(4)} \textbf{\add{Model variations}} \add{(2x settings): }
    \begin{itemize}
      \item   \underline{\add{GPT-3.5-turbo}}:  \add{this is the most recent version of GPT-3.5, known as gpt-3.5-turbo-1106.}
      \item   \underline{\add{GPT-4}}:  \add{this is the stable version of GPT-4, which points to gpt-4-0613.}
    \end{itemize}
\end{itemize}

\add{This creates a total of 24 prompt variants. We additionally ``union'' the best results from the 24 prompts, by counting a test case as correct as long as one of the 24 prompts can correctly predict the ground-truth formula, which we will report as \underline{GPT-Union (best-of-24-prompts)}.}


\etitle{Weak Supervision}. We compare with a simple version of the our method that uses only weak-supervision, with two sheets being ``similar'' if they pass our hypothesis-tests (Section~\ref{sec:traindatagen}). Like Mondrian, we use the formula found from the reference-sheet that is closest to the target-cell as the predicted formula.



\etitle{\sys}. This is our proposed method  in Section~\ref{sec:method}. 
In our experiments, use a view window of $100 \times 10$ (100 rows and 10 columns). 
The dimensionality of our embedding for the coarse-grained model is $896$, while that of the fine-grained model is $16000$ ($16$ dimensions per cell, times $10 \times 100$ cells in the view-window). 

\subsection{Quality Comparisons}



\input{tables/timestamp_res}

\begin{figure*}

	\includegraphics[width=1\textwidth]{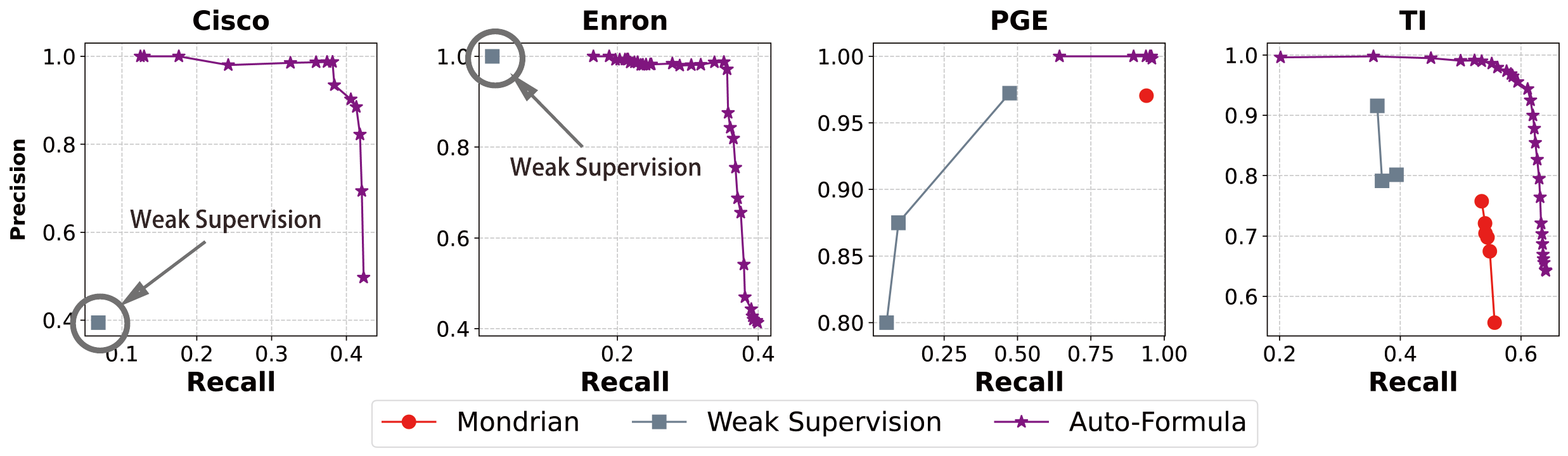}
	\caption{Quality comparisons using PR curves, on all test cases from 4 test corpora (Cisco/Enron/PGE/TI). Mondrian times out on two corpora (Cisco and Enron), and are thus not shown on two of these figures.}
	\label{fig:quality_comparasion}
\end{figure*}

\stitle{Comparison with Mondrian and Weak-Supervision.} 
We show the key precision/recall numbers in Table~\ref{tab:quality_timestamp}, and the corresponding PR-curves in Figure~\ref{fig:quality_comparasion}. These results are tested on all 2932 sampled formulas shown in Table~\ref{tab:datainfo}. 
\sys substantially outperforms other methods in all four test corpus (\code{Cisco/Enron/PGE/TI}).

\iftoggle{fullversion}{\input{tables/overall_res}}



\sys produces high-precision predictions (over 0.9 for the top-1 prediction), consistently across all 4 test corpora. We believe this is crucial to ensure good user experience, because users will likely find persistently incorrect predictions annoying. 
The recall of \sys, however, varies substantially (from 0.3 to 0.9) across different test corpora. We believe this is influenced by the characteristics of the underlying spreadsheets -- for certain test corpus (e.g., ``\code{Cisco}''), many of the underlying spreadsheets are ``singletons'', with a unique design pattern and no ``similar-sheets'' from the corpus for us to learn from, which limits the ``best possible recall'' of any similar-sheet-based methods.


Mondrian produces lower precision and recall, and we note that it times-out on 2 out of 4 test corpora after running for 1 week, because it uses a variant of agglomerative clustering that is cubic in the number of spreadsheets.



Weak-supervision produces good precision, but low recall. This is expected because weak-supervision  employs rigid  name-based rules to ensure high precision, which will miss out on similar-sheets that are named differently, thus lowering its recall. 


 
\stitle{Comparison with SpreadsheetCoder.} Because we need to manually trigger the recommendation feature in Google Sheets, for this test we evaluate using 180 randomly-sampled test formulas. 

Table~\ref{tab:quality_sampled} shows that SpreadsheetCoder produces substantially lower accuracy. This is not entirely surprising, because it is really difficult (even for humans) to infer a desired formula from only the natural language context (e.g., Figure~\ref{fig:formula-ex}), especially for complex formulas. 


\input{tables/gpt_28}

\stitle{Comparison with GPT. }Table~\ref{tab:gpt_variants} shows detailed results using 24 different prompt strategies presented. Among all prompts, the RAG-based method has the best F1 of 0.25, which
however is still substantially lower than our  method with over 0.9 F1.

In Table~\ref{tab:quality_sampled}, we add a method marked as ``GPT-union: best-of-24-prompt'', where we ``union'' all 24 prompts, and mark a case as ``correct'' for GPT as long as one prompt can get the case right (this is optimistic since without ground-truth, we don't know which prompt is the best beforehand). Even with the optimistic ``union'', GPT's result is only around 0.5. We carefully analyzed GPT results, and find the low performance of GPT unsurprising, because: (1) the
desired formula is often complex with multiple functions and parameters, which sometimes are hard even for humans
to guess (e.g., the formula in Figure 1); (2) spreadsheet formulas can often involve multiple tables, but large spreadsheet
context with multiple tables often cannot fit in GPT’s 4096-token context, leading to incorrect predictions; (3) our task
requires GPT to predict formula template and parameters correctly, which is challenging, like discussed above; (4) finally, we believe GPT has not seen many spreadsheets (.XLSX files) in its pre-training, making it not best suited for
spreadsheet formula predictions.

\input{tables/overall_res_google}




\begin{figure*}
    \begin{minipage}[t]{0.35\linewidth}
	\includegraphics[width=1\textwidth]{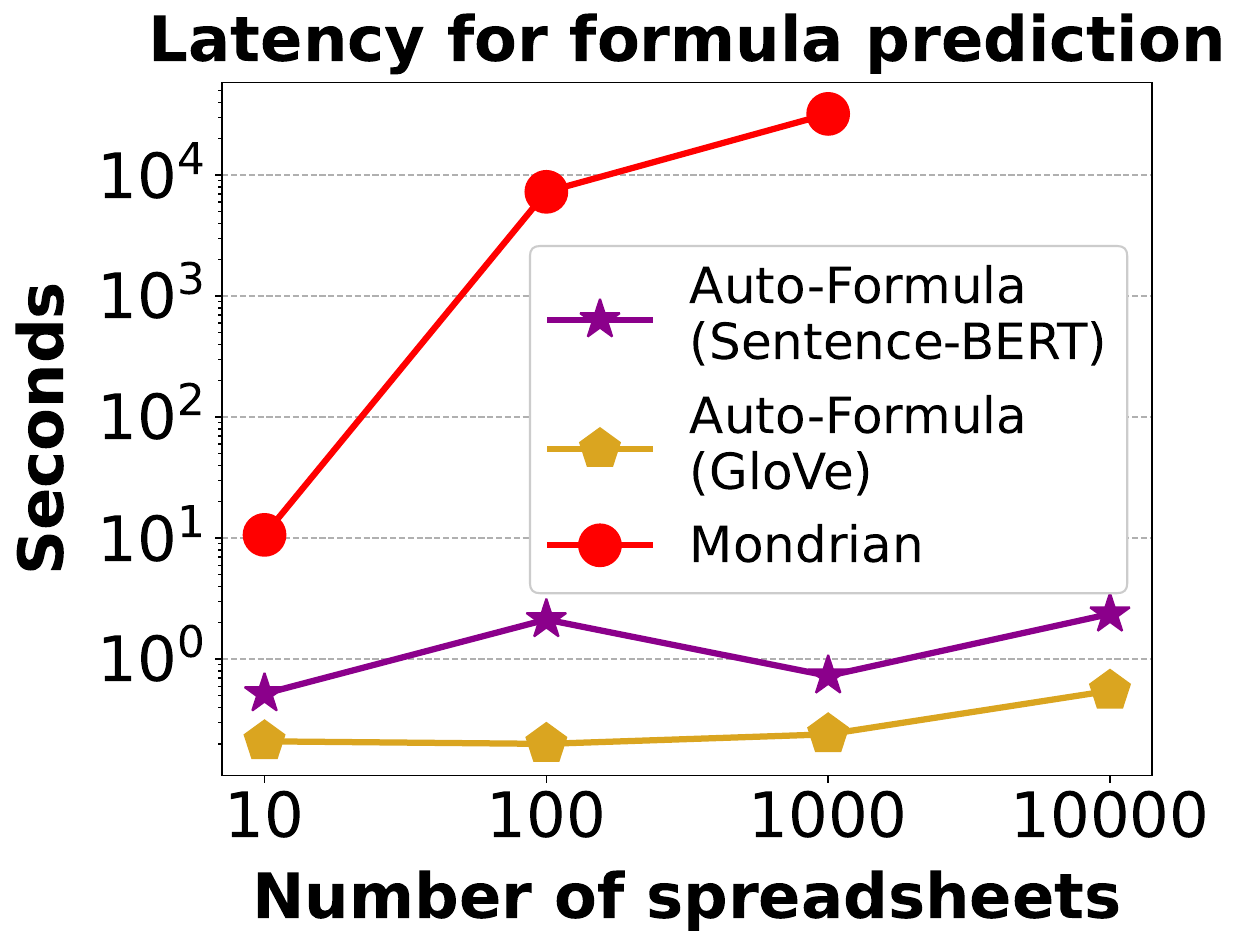}
    \vspace{-6mm}
	\caption{Scalability comparisons: as we increase the number of ``reference'' spreadsheets in the same enterprise.}
	\label{fig:latency}
 \end{minipage} \hspace{1em}
    \begin{minipage}[t]{0.55\linewidth}
 \includegraphics[width=1\columnwidth]{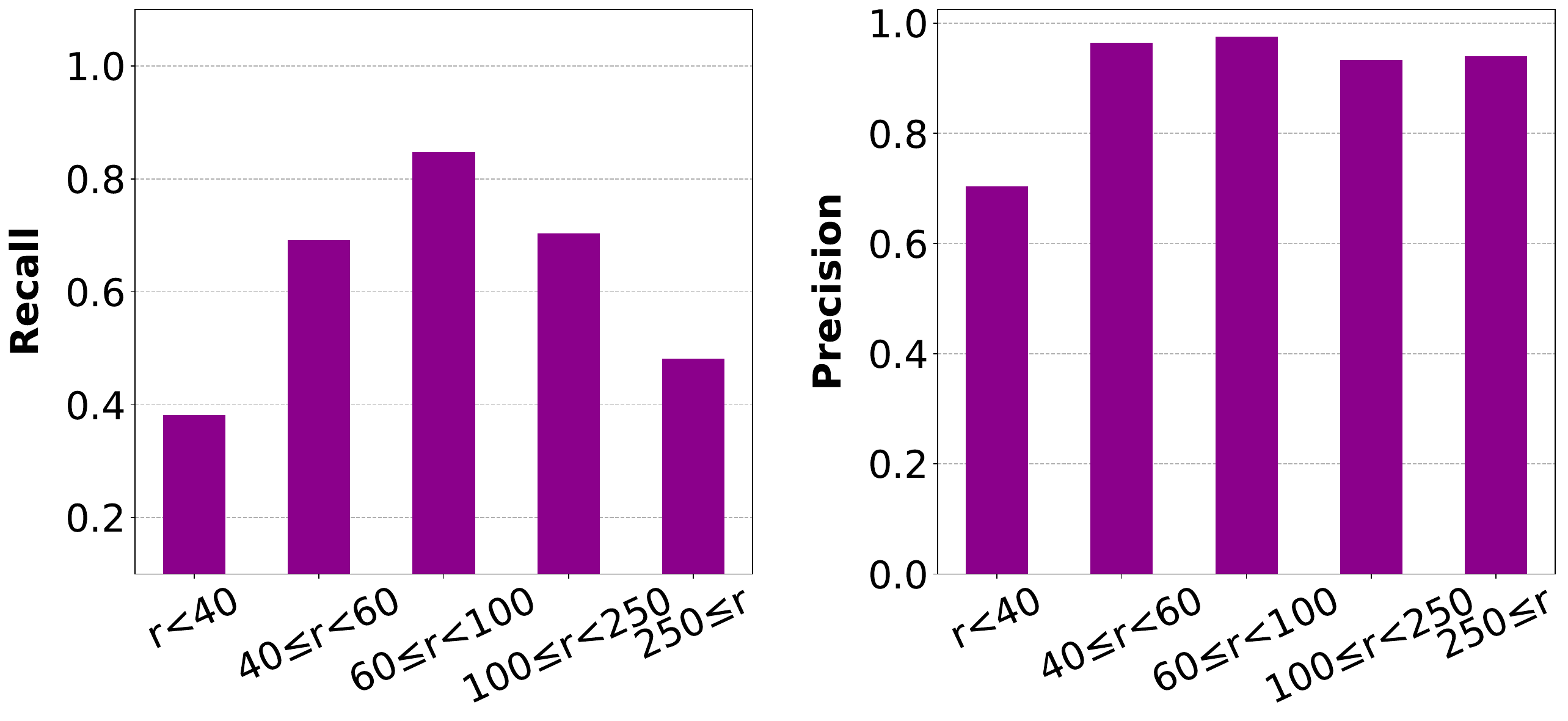}
     \vspace{-6mm}
    	\caption{Varying number of rows: we bucketize test cases based on the number of rows in the target sheet ($r$), which are shown on the x-axis.}
    	\label{fig:sheetsize}
     \end{minipage}
    	\vspace{-5mm}
\end{figure*}

\begin{figure*}
    \begin{minipage}[t]{0.45\linewidth}
 	\includegraphics[width=0.8\linewidth]{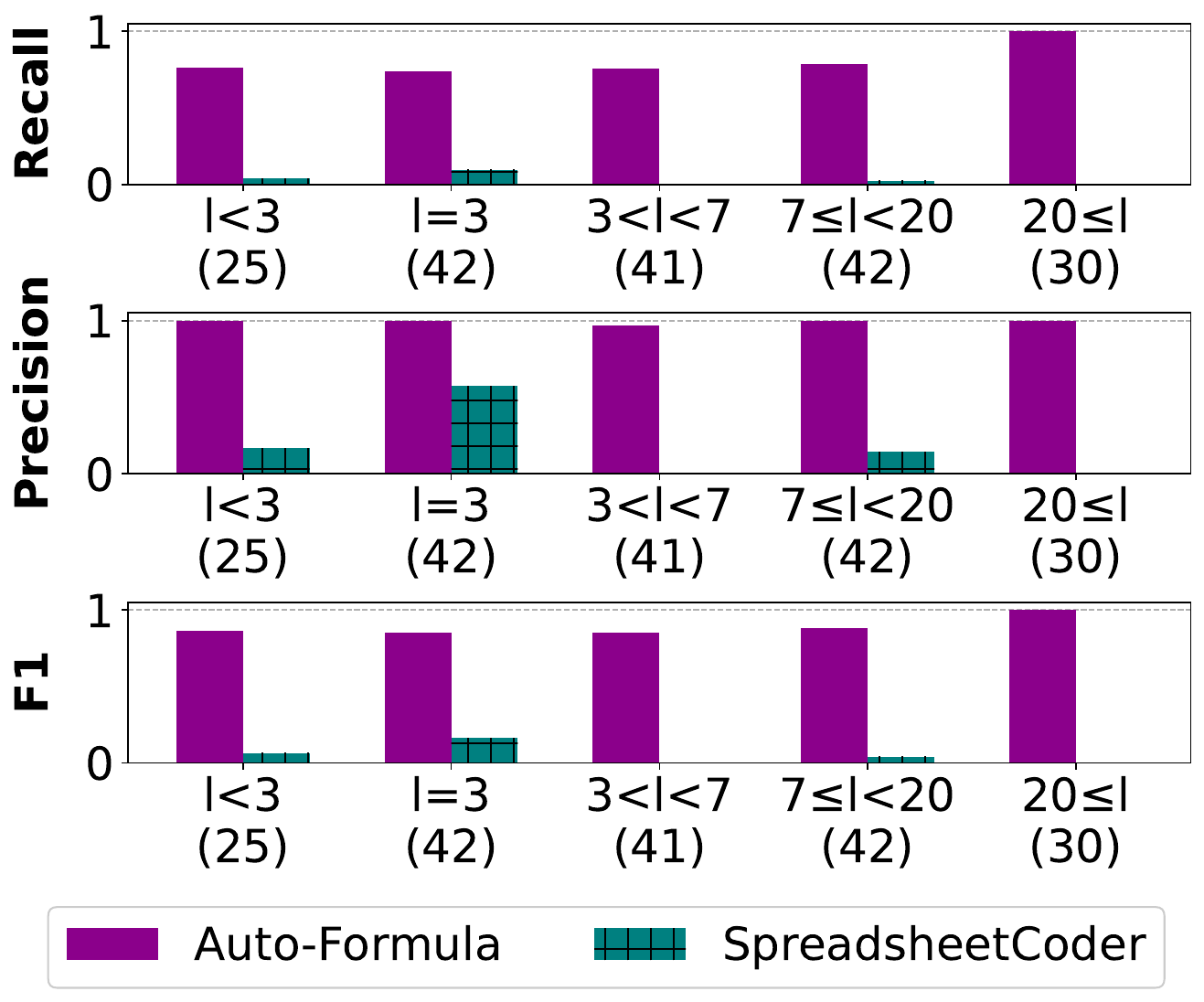}
	\vspace{-2mm}
	\caption{Quality comparisons: formulas are bucketized based on formula lengths (complexity).}
	\label{fig:formula_len}
    \end{minipage} \hspace{1em}
    \begin{minipage}[t]{0.45\linewidth}
 	\includegraphics[width=0.8\linewidth]{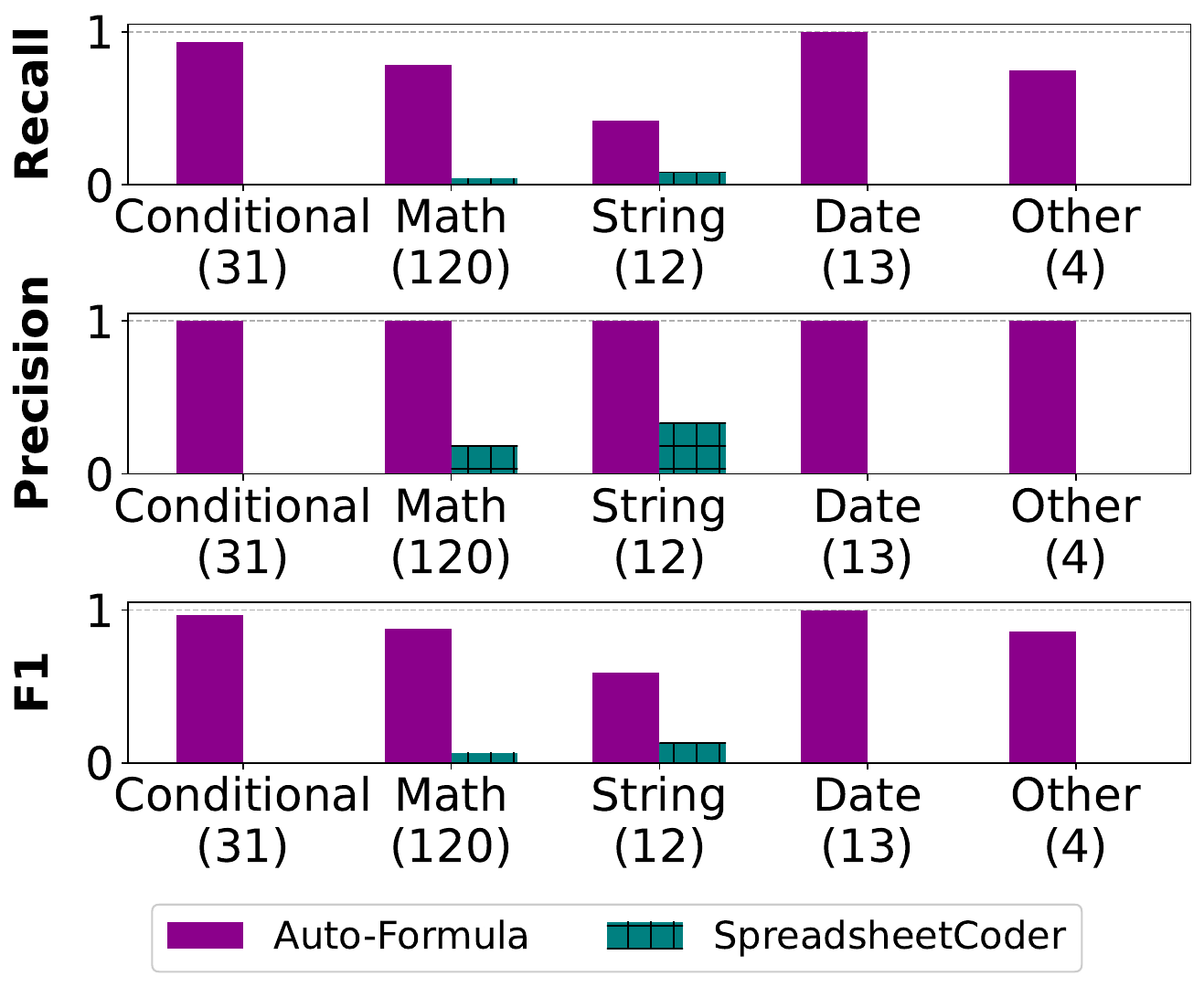}
 \vspace{-2mm}
	\caption{Quality comparisons: formulas are bucketized based on formula types (math, string, etc.).}
	\label{fig:formula_type}
    \end{minipage}
    \vspace{-2mm}
\end{figure*}


\subsection{Efficiency Comparisons} \label{sec:latency}
We evaluate the latency and scalability of different methods on real spreadsheet data.
Recall that the formula-recommendation problem requires interactive response time, because we need to make a prediction right when users select a target spreadsheet cell, so that they can verify/accept the suggested formula. 
Therefore, for each method, we differentiate between two types of running time: (1) pre-processing (offline), which is the time it takes to process the entire spreadsheet in preparation for online predictions, and (2) formula prediction (online),  which is the time it takes to make an online prediction, after users selecting a target cell. 

Figure~\ref{fig:latency} shows the latency of the crucial part of online prediction, where we vary the number of underlying spreadsheets available in an organization, from 10 to 10000.

We observed that \sys is orders of magnitude faster than Mondrian (which takes over 3 hours to process 1000 spreadsheets, and timed out after 1 week for 10000 spreadsheets). \sys is substantially more efficient, because we represent spreadsheet-regions as dense vectors, for which approximate nearest-neighbors (ANN) can be found efficiently leveraging recent advances in this area (we build ANN indexes using Faiss~\cite{faiss}). 
In comparison, Mondrian uses custom-made graph-matching that is not easy to index, and its clustering step is similar to agglomerative clustering with cubic complexity, making it hard to scale to large spreadsheet collections. 

Between two \sys variants, with GloVe and Sentence-BERT embedding, we find Sentence-BERT to be more expensive, though both are acceptable, with sub-second latency even with 10000 spreadsheets.

For offline pre-processing, we report the average latency to process one spreadsheet, for Mondrian, \sys-with-Glove and \sys-with-Sentence-BERT as $2051.05$, $55.83$ and $0.88$ seconds, respectively.

\subsection{Sensitivity Analysis}
\label{sec:sensitivity}

{

}

    
\begin{figure*}[t!]

     
    \includegraphics[width=0.95\textwidth]{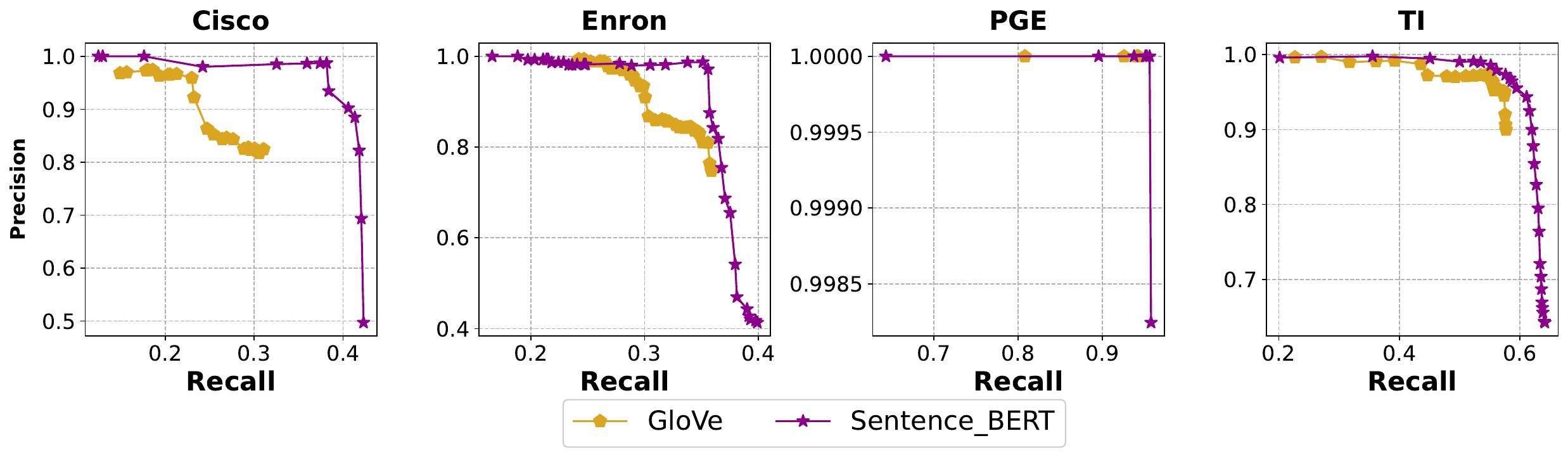}
	\caption{PR-curves: Sensitivity to embedding used (GloVe vs. Setence-BERT)}
	\label{fig:glove}

     \includegraphics[width=0.95\textwidth]{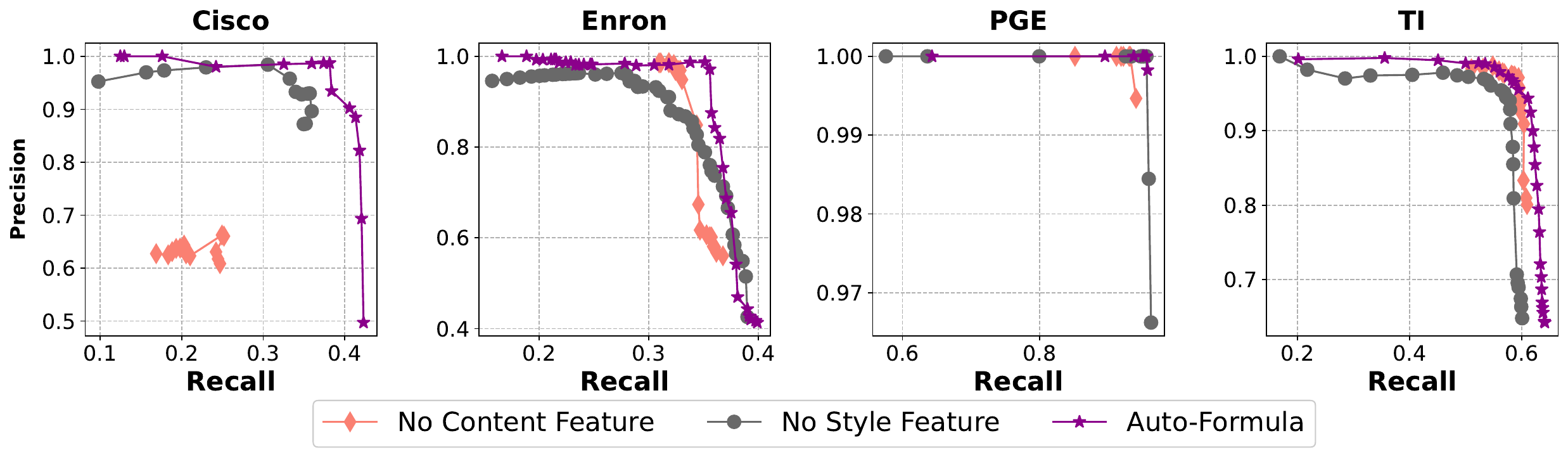}
	\caption{Ablation study: PR-curves without using content and style features.}
	\label{fig:feature}
    \includegraphics[width=0.95\textwidth]{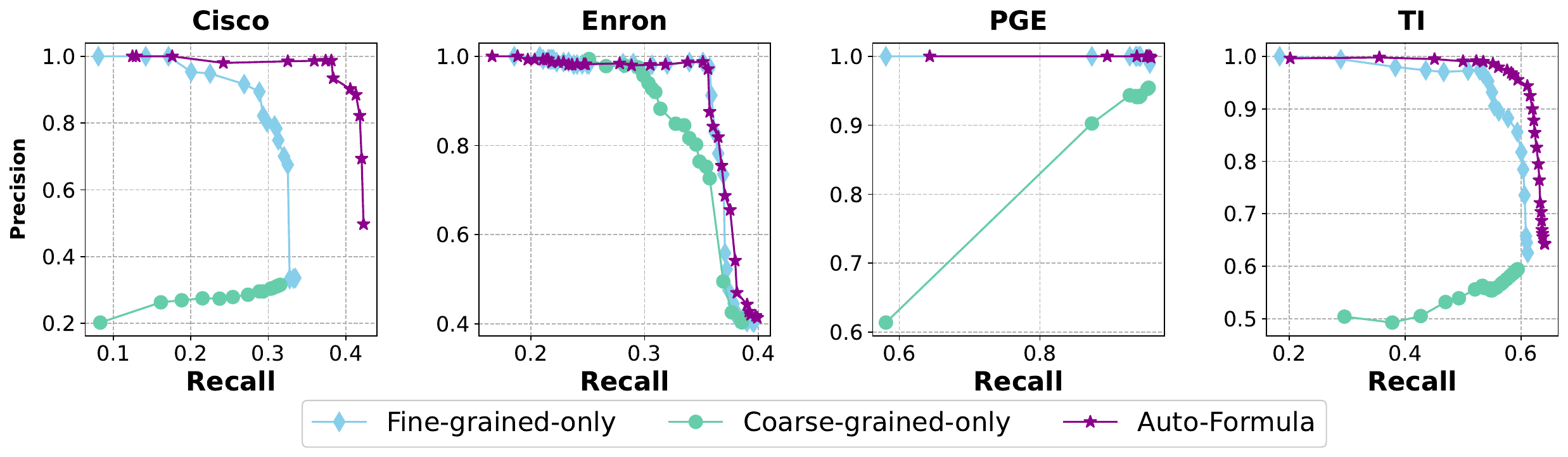}
	\caption{Ablation study: PR-curves without the separation of coarse-grained and fine-grained models.}
	\label{fig:only}
 

 \includegraphics[width=0.95\textwidth]{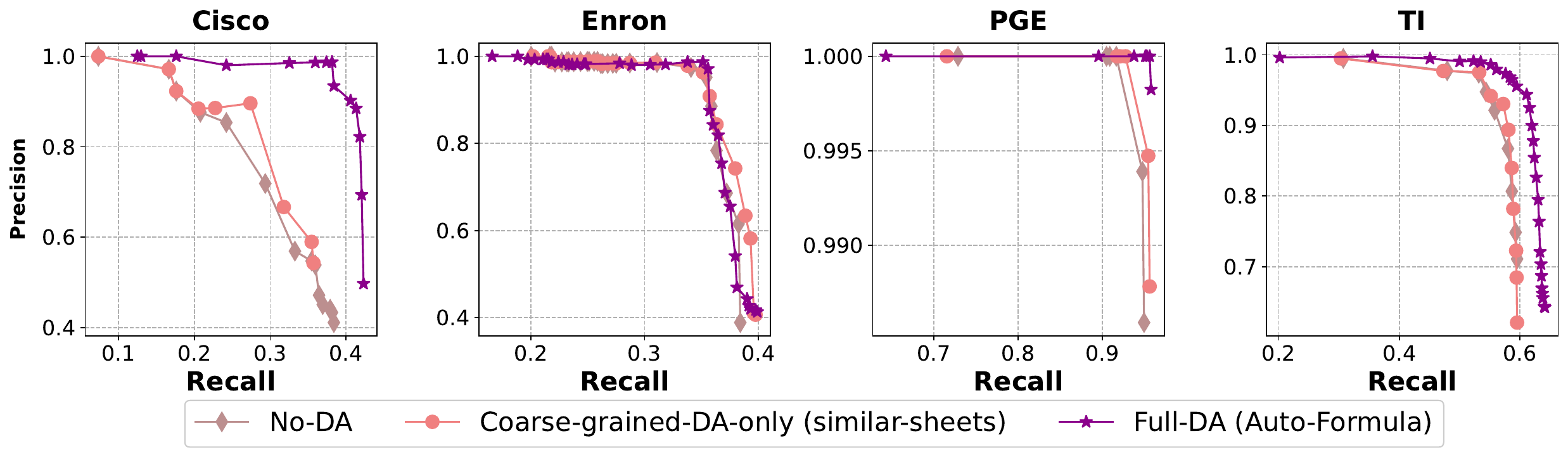}
	\caption{Effect of Data Augmentation (DA). Quality comparison between (1) No Data Augmentation; (2) With coarse-grained-DA-only; (3) with Full-DA (\sys), on all 4 test corpora.}
    \label{fig:da_ablation}
    
\end{figure*}

    \stitle{Sensitivity to sheet size}. To measure the impact of the size of the target spreadsheet on \sys, we bucketize the test cases by the number of rows, and report the resulting precision/recall in  Figure~\ref{fig:sheetsize}. For $recall$, there is a significant variation. This is because $recall$ is influenced by the factor of whether similar sheets exist. For $precision$, we observed that test cases with a smaller number of rows (less than $40$) have a lower precision, at $0.704$, while the precision for other test cases is around $0.95$. This occurs because when the sheets size are significantly smaller than the window size, the window is filled with a substantial amount of blank cells. Consequently, the model perceives these two windows to be more similar. For test cases where the sheets can fully occupy the window, \sys performs exceptionally well.
    

\stitle{Sensitivity to formula complexity}. 
Since longer formulas are naturally more difficult to predict, we define formula complexity as the number of nodes in its parsed abstract syntax tree, which corresponds to its length. We group formulas based on their lengths, and report the results in Figure~\ref{fig:formula_len}. The $recall$ of \sys in different groups is not significantly different, while the $precision$ is all close to $1$, indicating that \sys is not sensitive to the complexity of the varying formula. For SpreadsheetCoder, it performs  better on formulas with three or less nodes, suggesting that it is relatively more successful with simple formulas. 
    
\stitle{Sensitivity to formula types.} We also categorize the formulas into five types: "conditional" (with \code{IF-ELSE}), "math", "string", "date" and "other" and report our results in each category in Figure~\ref{fig:formula_type}. \sys is also not sensitive to formula types, except for type ``string'', where its recall  dips, suggesting that string-transformations are likely more ad-hoc in nature and more difficult to learn from similar sheets. For SpreadsheetCoder, we found that it performs better in simple math calculations (e.g., \code{SUM}).

    


\stitle{Sensitivity to embedding models.}  Since we use two alternative embedding models to obtain the content representation of spreadsheet cells, GloVe and Sentence-BERT, we study their impact in Figure~\ref{fig:glove}.
Overall, we observe that the two have similar quality, except on PGE, where Sentence-BERT has a slight advantage over GloVe.
Given that GloVe is shown to be noticeably faster than Sentence-BERT, 
we believe this presents a natural trade-off between quality (Sentence-BERT is slightly better) and efficiency (GloVe is more efficient), for practical applications.

\subsection{Ablation studies} 
\label{sec:ablation}

We performs ablation studies to understand the benefits of different components in \sys. 

\stitle{No content or style features}. In order to see the imporance of the content and sytle-based features in our spreadsheet representation, we remove the content features and style features, respectively, and report their effects in Figure~\ref{fig:feature}. It can be seen that both types of features are important, as removing either leads to a substantial drop in quality. 

\stitle{No separation of coarse-grained/fine-grained similarity.} Recall that in \sys, we create two variants of similarity models, one ``coarse-grained'' similarity for the detection of similar sheets, and another ``fine-grained'' similarity for the detection of similar regions.  To see the benefit of the separation, we create an ablation study in which we run the \sys end-to-end, using only the coarse-grained model or the fine-grained embedding model, and report the results in Figure~\ref{fig:only}. We can see that \sys outperforms both Coarse-grained-only and Fine-grained-only, with the gain over Coarse-grained-only being substantial, showing the need of fine-grained models to tell the subtle differences between two spreadsheet-regions that shifted only slightly, which however is crucial to correct prediction formula-templates and reference-ranges.


Compared to "Fine-grained-only," the performance of \sys is slightly but noticeably better, because it is more capable of detecting similar-sheets that have different rows and columns than the fine-grained model. We note that using coarse-grained models in similar-sheet detection has benefits beyond quality, because coarse-grained models use substantially smaller embedding vectors (892 vs. 16000 for fine-grained models), making it  orders of magnitude more efficient to store, index and query, using ANN-based indexing techniques.




{\stitle{No Data Augmentation (DA).}
Figure~\ref{fig:da_ablation} shows the results with and without data augmentation (DA). We observe a sizable drop in quality with No-DA, as augmentation (by removing rows/columns) allows the model to identify similar sheets/regions in a more robust manner. Coarse-grained-DA-only (for similar sheets) yields a similar drop in quality on average.
}



%% file: tables/tableinfo.tex
\begin{table}[!t]
\vspace{-3mm}
\caption{Statistics of test data.}
\vspace{-4mm}
\label{tab:quality}
 \resizebox{0.7\columnwidth}{!}{
\begin{tabular}{|c|c|c|c|c|c|}
\hline
               &  All        & PGE   & Cisco  & TI     & Enron   \\ \hline
\# of workbooks   & 12,750     & 459   & 213    & 1,549   & 10,529   \\ \hline
\# of sheets      & 51,037    & 1,214  & 682    & 4,200   & 44,941   \\ \hline
\# of formulas    & 3,056,810 & 45,268 & 357,018 & 258,403 & 2,396,121 \\ \hline
\# of \code{test formulas (Random)}     & 3,815      & 1,000  & 923    & 1,000   & 892     \\ \hline
\# of \code{test formulas (Timestamps)}     & 2,932      & 594  & 409    & 1,260   & 669     \\ \hline
\end{tabular}

 }
\label{tab:datainfo}
\vspace{-1em}
\end{table}

%% file: tables/timestamp_res.tex
\begin{table*}[t!]
\caption{Quality comparisons of all test cases from 4 test corpora (Cisco/Enron/PGE/TI), where we report Recall (R), Precision (P), and F1. The leftmost ``Overall Average'' column reports  average results on these 4 corpora. Figure~\ref{fig:quality_comparasion} shows the corresponding PR curves of the numbers reported here.}
\vspace{-2mm}
\label{tab:quality_timestamp}
\resizebox{1\textwidth}{!}{

\begin{tabular}{c||ccc||ccc|ccc|ccc|ccc}
\hline
                   & \multicolumn{3}{c||}{Overall Average}               & \multicolumn{3}{c|}{Cisco}                         & \multicolumn{3}{c|}{Enron}                      & \multicolumn{3}{c|}{PGE}                          & \multicolumn{3}{c}{TI}                    \\ \hline
Metric             & R         & P      & F1             & R         & P      & F1          & R         & P      & F1             & R         & P  & F1             & R         & P  & F1             \\ \hline
Auto-Formula       &\textbf{0.54}&\textbf{0.99}&\textbf{0.70}                 & \textbf{0.36 } & \textbf{0.99 } & \textbf{0.53}    & \textbf{0.34}&0.99&\textbf{0.50}    &\textbf{0.94}&\textbf{1}&\textbf{0.97}    &\textbf{0.54}&\textbf{0.99}&\textbf{0.69}\\
Mondrian           &0.39&0.43&0.48   & \multicolumn{3}{c|}{[Time Out]}   & \multicolumn{3}{c|}{[Time Out]}   & 0.93& 0.97& 0.95   &0.54& 0.76& 0.63         \\
Weak Supervision &0.24&0.78&0.33     & 0.07& 0.39& 0.12    & 0.02& \textbf{1}& 0.04    & 0.47& 0.97& 0.64       & 0.39& 0.75& 0.52   \\ \hline
\end{tabular}
}


\end{table*}

%% file: tables/overall_res.tex
\begin{table*}[t!]
\caption{Quality comparisons (using Random Split) on all test cases from 4 test corpora (Cisco/Enron/PGE/TI), where we report Recall (R), Precision (P), and F1. The leftmost ``Overall Average'' column reports  average results on these 4 corpora. This is the ``random-split'' version of the test in Table~\ref{tab:quality_timestamp} (which uses ``timestamp-split'').}
\vspace{-2mm}
\label{tab:quality_random}
\resizebox{1.0\textwidth}{!}{

\begin{tabular}{c||ccc||ccc|ccc|ccc|ccc}
\hline
                   & \multicolumn{3}{c||}{Overall Average}               & \multicolumn{3}{c|}{Cisco}                         & \multicolumn{3}{c|}{Enron}                      & \multicolumn{3}{c|}{PGE}                          & \multicolumn{3}{c}{TI}                    \\ \hline
Metric             & R         & P      & F1             & R         & P      & F1          & R         & P      & F1             & R         & P  & F1             & R         & P  & F1             \\ \hline
Auto-Formula       &\textbf{0.60}&\textbf{0.89}&\textbf{0.71}                 & \textbf{0.42 } & \textbf{0.86 } & \textbf{0.56}    & \textbf{0.62}&0.91&\textbf{0.74}    &\textbf{0.93}&0.98&\textbf{0.96}    &\textbf{0.45}&\textbf{0.82}&\textbf{0.59}\\
Mondrian           &0.33&0.80&0.47   & \multicolumn{3}{c|}{[Time Out]}   & \multicolumn{3}{c|}{[Time Out]}   & 0.90& 0.97& 0.94   &0.44& 0.64& 0.52         \\
Weak Supervision &0.39&0.83&0.48     & 0.15& 0.62& 0.24    & 0.16& \textbf{1}& 0.28    & 0.90& \textbf{0.99}& 0.94       & 0.35& 0.72& 0.47   \\ \hline
\end{tabular}
}

\end{table*}

%% file: tables/gpt_28.tex
\begin{table}[]
\caption{``GPT-Union'' results using 24 prompt engineering variants.}
\label{tab:gpt_variants}
\resizebox{0.75 \columnwidth}{!}{
\begin{tabular}{|cccc|c|c|c|}
\hline
\multicolumn{1}{|c|}{\begin{tabular}[c]{@{}c@{}} Example selection \end{tabular}}                       & \multicolumn{1}{c|}{\begin{tabular}[c]{@{}c@{}}Chain of \\ Thought\end{tabular}}             & \multicolumn{1}{c|}{\begin{tabular}[c]{@{}c@{}}Table \\ Regions\end{tabular}} & \begin{tabular}[c]{@{}c@{}}Model \\ Variations\end{tabular} & Recall & Precision & F1    \\ \hline
\multicolumn{1}{|c|}{\multirow{8}{*}{Zero-shot}}                                                           & \multicolumn{1}{c|}{\multirow{4}{*}{With COT}}                                               & \multicolumn{1}{c|}{\multirow{2}{*}{Precise-table}}                           & GPT-3.5                                                     & 0      & 0         & 0     \\ \cline{4-7} 
\multicolumn{1}{|c|}{}                                                                                     & \multicolumn{1}{c|}{}                                                                        & \multicolumn{1}{c|}{}                                                         & GPT-4                                                       & 0.033  & 0.034     & 0.033 \\ \cline{3-7} 
\multicolumn{1}{|c|}{}                                                                                     & \multicolumn{1}{c|}{}                                                                        & \multicolumn{1}{c|}{\multirow{2}{*}{Large-sheet}}                             & GPT-3.5                                                     & 0      & 0         & 0     \\ \cline{4-7} 
\multicolumn{1}{|c|}{}                                                                                     & \multicolumn{1}{c|}{}                                                                        & \multicolumn{1}{c|}{}                                                         & GPT-4                                                       & 0.033  & 0.033     & 0.034 \\ \cline{2-7} 
\multicolumn{1}{|c|}{}                                                                                     & \multicolumn{1}{c|}{\multirow{4}{*}{\begin{tabular}[c]{@{}c@{}}Without \\ COT\end{tabular}}} & \multicolumn{1}{c|}{\multirow{2}{*}{Precise-table}}                           & GPT-3.5                                                     & 0.011  & 0.011     & 0.011 \\ \cline{4-7} 
\multicolumn{1}{|c|}{}                                                                                     & \multicolumn{1}{c|}{}                                                                        & \multicolumn{1}{c|}{}                                                         & GPT-4                                                       & 0.044  & 0.045     & 0.044 \\ \cline{3-7} 
\multicolumn{1}{|c|}{}                                                                                     & \multicolumn{1}{c|}{}                                                                        & \multicolumn{1}{c|}{\multirow{2}{*}{Large-sheet}}                             & GPT-3.5                                                     & 0      & 0         & 0     \\ \cline{4-7} 
\multicolumn{1}{|c|}{}                                                                                     & \multicolumn{1}{c|}{}                                                                        & \multicolumn{1}{c|}{}                                                         & GPT-4                                                       & 0.039  & 0.039     & 0.039 \\ \hline
\multicolumn{1}{|c|}{\multirow{8}{*}{\begin{tabular}[c]{@{}c@{}}Few-shot, \\ common-formula\end{tabular}}} & \multicolumn{1}{c|}{\multirow{4}{*}{With COT}}                                               & \multicolumn{1}{c|}{\multirow{2}{*}{Precise-table}}                           & GPT-3.5                                                     & 0.006  & 0.006     & 0.006 \\ \cline{4-7} 
\multicolumn{1}{|c|}{}                                                                                     & \multicolumn{1}{c|}{}                                                                        & \multicolumn{1}{c|}{}                                                         & GPT-4                                                       & 0.044  & 0.047     & 0.045 \\ \cline{3-7} 
\multicolumn{1}{|c|}{}                                                                                     & \multicolumn{1}{c|}{}                                                                        & \multicolumn{1}{c|}{\multirow{2}{*}{Large-sheet}}                             & GPT-3.5                                                     & 0      & 0         & 0     \\ \cline{4-7} 
\multicolumn{1}{|c|}{}                                                                                     & \multicolumn{1}{c|}{}                                                                        & \multicolumn{1}{c|}{}                                                         & GPT-4                                                       & 0.017  & 0.018     & 0.017 \\ \cline{2-7} 
\multicolumn{1}{|c|}{}                                                                                     & \multicolumn{1}{c|}{\multirow{4}{*}{\begin{tabular}[c]{@{}c@{}}Without \\ COT\end{tabular}}} & \multicolumn{1}{c|}{\multirow{2}{*}{Precise-table}}                           & GPT-3.5                                                     & 0.006  & 0.006     & 0.006 \\ \cline{4-7} 
\multicolumn{1}{|c|}{}                                                                                     & \multicolumn{1}{c|}{}                                                                        & \multicolumn{1}{c|}{}                                                         & GPT-4                                                       & 0.039  & 0.039     & 0.039 \\ \cline{3-7} 
\multicolumn{1}{|c|}{}                                                                                     & \multicolumn{1}{c|}{}                                                                        & \multicolumn{1}{c|}{\multirow{2}{*}{Large-sheet}}                             & GPT-3.5                                                     & 0      & 0         & 0     \\ \cline{4-7} 
\multicolumn{1}{|c|}{}                                                                                     & \multicolumn{1}{c|}{}                                                                        & \multicolumn{1}{c|}{}                                                         & GPT-4                                                       & 0.028  & 0.028     & 0.028 \\ \hline
\multicolumn{1}{|c|}{\multirow{8}{*}{\begin{tabular}[c]{@{}c@{}}Few-shot, \\ RAG-formulas\end{tabular}}}   & \multicolumn{1}{c|}{\multirow{4}{*}{With COT}}                                               & \multicolumn{1}{c|}{\multirow{2}{*}{Precise-table}}                           & GPT-3.5                                                     & 0.206  & 0.211     & 0.208 \\ \cline{4-7} 
\multicolumn{1}{|c|}{}                                                                                     & \multicolumn{1}{c|}{}                                                                        & \multicolumn{1}{c|}{}                                                         & GPT-4                                                       & 0.233  & 0.235     & 0.234 \\ \cline{3-7} 
\multicolumn{1}{|c|}{}                                                                                     & \multicolumn{1}{c|}{}                                                                        & \multicolumn{1}{c|}{\multirow{2}{*}{Large-sheet}}                             & GPT-3.5                                                     & 0.239  & 0.242     & 0.24  \\ \cline{4-7} 
\multicolumn{1}{|c|}{}                                                                                     & \multicolumn{1}{c|}{}                                                                        & \multicolumn{1}{c|}{}                                                         & GPT-4                                                       & 0.172  & 0.174     & 0.173 \\ \cline{2-7} 
\multicolumn{1}{|c|}{}                                                                                     & \multicolumn{1}{c|}{\multirow{4}{*}{\begin{tabular}[c]{@{}c@{}}Without \\ COT\end{tabular}}} & \multicolumn{1}{c|}{\multirow{2}{*}{Precise-table}}                           & GPT-3.5                                                     & 0.256  & 0.263     & 0.259 \\ \cline{4-7} 
\multicolumn{1}{|c|}{}                                                                                     & \multicolumn{1}{c|}{}                                                                        & \multicolumn{1}{c|}{}                                                         & GPT-4                                                       & 0.239  & 0.24      & 0.239 \\ \cline{3-7} 
\multicolumn{1}{|c|}{}                                                                                     & \multicolumn{1}{c|}{}                                                                        & \multicolumn{1}{c|}{\multirow{2}{*}{Large-sheet}}                             & GPT-3.5                                                     & 0.244  & 0.249     & 0.246 \\ \cline{4-7} 
\multicolumn{1}{|c|}{}                                                                                     & \multicolumn{1}{c|}{}                                                                        & \multicolumn{1}{c|}{}                                                         & GPT-4                                                       & 0.161  & 0.162     & 0.161 \\ \hline
\multicolumn{4}{|c|}{\textbf{GPT-union (best-of-24-prompts)}}                                                                                                                                                                                                                                                                                                                             & 0.461  & 0.461     & 0.461 \\ \hline
\end{tabular}
}
\end{table}

%% file: tables/overall_res_google.tex
\begin{table}[!t]
\caption{Quality comparison with SpreadsheetCoder and GPT, on a sampled subset of 180 formulas.}
\vspace{-1em}
\label{tab:quality_sampled}
 \resizebox{0.6 \columnwidth}{!}{
\begin{tabular}{cccc}
\hline
                   & Recall       & Precision     & F1             \\ \hline
Auto-Formula       & \textbf{0.8} & \textbf{0.993} & \textbf{0.886} \\
SpreadsheetCoder & 0.039        & 0.171         & 0.064          \\ 
GPT-union (best-of-24-prompts)  & 0.461      & 0.461            & 0.461              \\ \hline    
\end{tabular}
 }
\end{table}

%% file: conclusions.tex
\section{Conclusions and Future Work}

In this paper, we study the problem of formula recommendation, where we develop a new approach to accurately predict formulas, by learning and adapting formulas from similar-sheets.  Future directions include extending the similar-sheet primitive to enable other spreadsheet applications, such as content auto-filling, table error detection, and security-related use cases.


